\documentclass[useAMS,usenatbib,iop,numberedappendix]{mn2e} 

\usepackage{amsmath,amsfonts,amssymb}
\usepackage{makeidx}
\usepackage{graphicx}
\usepackage{epstopdf}
\usepackage[colorlinks=false]{hyperref}
\usepackage{multirow}
\usepackage{rotating}
\usepackage{color}
\usepackage{tablefootnote}
\usepackage{lscape}
\usepackage{subfigure}
\usepackage{tablefootnote}
\usepackage[T1]{fontenc}
\usepackage{aecompl}
 
\definecolor{ored}{rgb}{1.00,0.27,0.00}

\newcommand{\Munich}{$^{1}$}
\newcommand{\ExcellenceCluster}{$^{2}$}
\newcommand{\MPE}{$^{3}$}
\newcommand{\MIT}{$^{4}$}
\newcommand{\CfA}{$^{5}$}
\newcommand{\Harvard}{$^{6}$}
\newcommand{\FNAL}{$^{7}$}
\newcommand{\KICPChicago}{$^{8}$}
\newcommand{\AAUChicago}{$^{9}$}
\newcommand{\PhysicsUChicago}{$^{10}$}
\newcommand{\ANL}{$^{11}$}
\newcommand{\Miss}{$^{12}$}
\newcommand{\UFlorida}{$^{13}$}
\newcommand{\Melbourne}{$^{14}$}
\newcommand{\UHawaii}{$^{15}$}
\newcommand{\UCDavis}{$^{16}$}
\newcommand{\IGPP}{$^{17}$}
\newcommand{\KoreaASSI}{$^{18}$}
\newcommand{\AIfA}{$^{19}$}
\newcommand{\CTIO}{$^{20}$}

\author[Chiu et al.]{
I.~Chiu\Munich$^,$\ExcellenceCluster,
J.~Mohr\Munich$^,$\ExcellenceCluster$^,$\MPE,
M.~McDonald\MIT,
S.~Bocquet\Munich$^,$\ExcellenceCluster,
M.~L.~N.~Ashby\CfA,
\newauthor 
M.~Bayliss\Harvard$^,$\CfA,
B.~A.~Benson\FNAL$^,$\KICPChicago$^,$\AAUChicago,
L.~E.~Bleem\KICPChicago$^,$\PhysicsUChicago$^,$\ANL,
M.~Brodwin\Miss,
S.~Desai\Munich$^,$\ExcellenceCluster,
\newauthor 
J.~P.~Dietrich\Munich$^,$\ExcellenceCluster,
W.~R.~Forman\CfA,
C.~Gangkofner\Munich$^,$\ExcellenceCluster,
A.~H.~Gonzalez\UFlorida,
C.~Hennig\Munich$^,$\ExcellenceCluster,
\newauthor 
J.~Liu\Munich$^,$\ExcellenceCluster,
C.~L.~Reichardt\Melbourne, 
A.~Saro\Munich$^,$\ExcellenceCluster,
B.~Stalder\CfA$^,$\UHawaii,
S.~A.~Stanford\UCDavis$^,$\IGPP,
J.~Song \KoreaASSI,
\newauthor 
T.~Schrabback\AIfA,
R.~\v Suhada\Munich,
V.~Strazzullo\Munich,
A.~Zenteno\Munich$^,$\CTIO
\\
\Munich Department of Physics, Ludwig-Maximilians-Universit\"{a}t, Scheinerstr.\ 1, 81679 M\"{u}nchen, Germany \\
\ExcellenceCluster Excellence Cluster Universe, Boltzmannstr.\ 2, 85748 Garching, Germany \\
\MPE Max-Planck-Institut f\"{u}r extraterrestrische Physik, Giessenbachstr.\ 85748 Garching, Germany \\
\MIT Kavli Institute for Astrophysics and Space Research, Massachusetts Institute of Technology, 77 Massachusetts Avenue, Cambridge, MA 02139 \\
\CfA Harvard-Smithsonian Center for Astrophysics, 60 Garden Street, Cambridge, MA 02138 \\
\Harvard Department of Physics, Harvard University, 17 Oxford Street, Cambridge, MA 02138 \\
\FNAL Fermi National Accelerator Laboratory, Batavia, IL 60510-0500 \\
\KICPChicago Kavli Institute for Cosmological Physics, University of Chicago, 5640 South Ellis Avenue, Chicago, IL 60637 \\
\AAUChicago Department of Astronomy and Astrophysics, University of Chicago, 5640 South Ellis Avenue, Chicago, IL 60637 \\
\PhysicsUChicago Department of Physics, University of Chicago, 5640 South Ellis Avenue, Chicago, IL 60637 \\
\ANL Argonne National Laboratory, 9700 S. Cass Avenue, Argonne, IL, USA 60439 \\
\Miss Department of Physics and Astronomy, University of Missouri, 5110 Rockhill Road, Kansas City, MO 64110 \\
\UFlorida Department of Astronomy, University of Florida, Gainesville, FL 32611 \\
\Melbourne School of Physics, University of Melbourne, Parkville, VIC 3010, Australia \\
\UHawaii Institute for Astronomy, University of Hawaii at Manoa, Honolulu, HI 96822, USA \\
\UCDavis Department of Physics, University of California, One Shields Avenue, Davis, CA 95616 \\
\IGPP Institute of Geophysics and Planetary Physics, Lawrence Livermore National Laboratory, Livermore, CA 94550 \\
\KoreaASSI Korea Astronomy and Space Science Institute 776, Daedeokdae-ro, Yuseong-gu, Daejeon, Republic of Korea 305-348 \\
\AIfA Argelander-Institut f\"ur Astronomie, Auf dem H\"ugel 71, D-53121 Bonn, Germany \\
\CTIO Cerro Tololo Inter-American Observatory, Casilla 603, La Serena, Chile
}

\newcommand{\LCDM}{\ensuremath{\Lambda\mathrm{CDM}}}
\newcommand{\OmegaM}{\ensuremath{\Omega_{\mathrm{M}}}}
\newcommand{\OmegaL}{\ensuremath{\Omega_{\Lambda} } }

\newcommand{\Hnow}{\ensuremath{H_{0}}}

\newcommand{\CHANDRA}{\emph{Chandra}}
\newcommand{\XMMNEWTON}{XMM-\emph{Newton}}
\newcommand{\Spitzer}{\emph{Spitzer}}
\newcommand{\HST}{\emph{HST}}
\newcommand{\PLANCK}{\emph{Planck}}

\newcommand{\BHigh}{\ensuremath{\mathrm{b}_{\mathrm{H}}}}
\newcommand{\Fband}{\ensuremath{\mathrm{F606W}}}
\newcommand{\IBess}{\ensuremath{\mathrm{I}_{\mathrm{B}}}}
\newcommand{\ZGunn}{\ensuremath{\mathrm{z}_{\mathrm{G}}}}
\newcommand{\Cha}{\ensuremath{\left[3.6\right]}}
\newcommand{\Chb}{\ensuremath{\left[4.5\right]}}

\newcommand{\mstar}{\ensuremath{m^{\ast}}}
\newcommand{\Msun}{\ensuremath{M_{\odot}}}

\newcommand{\Mbest}{\ensuremath{M_{\star}^{\mathrm{best}}}}
\newcommand{\Mmed}{\ensuremath{M_{\star}^{\mathrm{med}}}}
\newcommand{\Mlo}{\ensuremath{M_{\star}^{\mathrm{lo}}}}
\newcommand{\Mhi}{\ensuremath{M_{\star}^{\mathrm{hi}}}}

\newcommand{\fgas}{\ensuremath{f_{\mathrm{ICM}}}}
\newcommand{\fstar}{\ensuremath{f_{\star}}}
\newcommand{\fb}{\ensuremath{f_{\mathrm{b}}}}
\newcommand{\fcold}{\ensuremath{f_{\mathrm{c}}}}

\newcommand{\Msz}{\ensuremath{M_{500}}}
\newcommand{\Mgas}{\ensuremath{M_{\mathrm{ICM}}}}
\newcommand{\Mstar}{\ensuremath{M_{\star}}}

\newcommand{\MBCG}{\ensuremath{M_{\star}^{\mathrm{BCG}}}}
\newcommand{\Mbaryon}{\ensuremath{M_{\mathrm{b}}}}
\newcommand{\Rfiveoo}{\ensuremath{R_{500}}}

\newcommand{\mzero}{\ensuremath{m_{0}}}
\newcommand{\Mzero}{\ensuremath{M_{0}}}
\newcommand{\percent}{\ensuremath{\%}}

\title[Baryon Content of Massive Galaxy Clusters]{Baryon Content of Massive Galaxy Clusters at $0.57<z<1.33$}

\begin{document}
\pdfpageheight 11.7in
\pdfpagewidth 8.3in

%
%

\maketitle 

%
%

\begin{abstract}
We study the stellar, Brightest Cluster Galaxy (BCG) and intracluster medium (ICM) masses of 14 South Pole Telescope (SPT) selected galaxy clusters with median redshift $z=0.9$ and mass $\Msz=6\times10^{14}\Msun$. We estimate stellar masses for each cluster and BCG using six photometric bands, the ICM mass using 
X-ray observations, and the virial masses using the SPT Sunyaev-Zel'dovich Effect signature.
At $z=0.9$ the BCG mass \MBCG\ constitutes $0.12\pm0.01$\percent\ of the halo mass for a $6\times10^{14}\Msun$ cluster, and this fraction falls as $\Msz^{-0.58\pm0.07}$. The cluster stellar mass function has a characteristic mass $\Mzero=10^{11.0\pm0.1}\Msun$, and the number of galaxies per unit mass in clusters is larger than in the field by a factor $1.65\pm0.20$.  
We combine our SPT sample with previously published samples at low redshift and correct to a common initial mass function and for systematic virial mass differences.  We then explore mass and redshift trends in the stellar fraction \fstar, the ICM fraction \fgas, the collapsed baryon fraction \fcold\ and the baryon fraction \fb.  At a pivot mass of $6\times10^{14}\Msun$ and redshift $z=0.9$, the characteristic values are \fstar=$1.1\pm0.1$\percent, \fgas=$9.6\pm0.5$\percent, \fcold=$10.7\pm1.1$\percent\ and \fb=$10.7\pm0.6$\percent. These fractions all vary with cluster mass at high significance, with higher mass clusters having lower \fstar\ and \fcold\ and higher \fgas\ and \fb. When accounting for a 15\percent\ systematic virial mass uncertainty, there is no statistically significant redshift trend at fixed mass.
Our results support the scenario where clusters grow through accretion from subclusters (higher \fstar, lower \fgas) and the field (lower \fstar, higher \fgas), balancing to keep \fstar\ and \fgas\ approximately constant since $z\sim0.9$.
\end{abstract}

%
%

\begin{keywords}
galaxy clusters - cosmology - galaxy evolution
\end{keywords}

%
%

\section{Introduction}
\label{sec:introduction}
The utility of galaxy clusters for cosmological parameter studies was recognized quite early \citep{frenk90,henry91,lilje92,white93a,white93b}, but the overwhelming evidence of widespread merging in the cluster population \citep{geller82,forman82,dressler88,mohr95} together with the high scatter in the X-ray luminosity--temperature relation \citep[e.g.,][]{fabian94a} left many with the impression that clusters were too complex and varied to ever be useful for cosmological studies. It was some time later that the first evidence that clusters exhibit significant regularity in their intracluster medium (ICM) properties appeared \citep{mohr97,arnaud99,cavaliere99,mohr99}; X-ray observations showed that clusters as a population exhibit a size--temperature scaling relation with $\approx10\percent$ scatter, a level of regularity comparable to that known in elliptical galaxies \citep[i.e.,][]{djorgovski87}. This regularity together with the emergence of evidence for cosmic acceleration \citep{riess98,perlmutter99} focused renewed interest in the use of galaxy clusters for precise cosmological studies \citep[e.g.,][]{haiman01}.  Moreover, the existence of low scatter, power law relations among cluster observables provided a useful tool to study the variation in cluster structure with mass and redshift.

Soon thereafter, the regularity seen in the X-ray properties of clusters was shown to exist also in the optical properties of clusters \citep[][hereafter L03]{lin03b}.  L03 carried out an X-ray and near-infrared (NIR) 2MASS $K$-band study of an ensemble of 27 nearby clusters, measuring the mass fraction of the stellar component inside the galaxies (\fstar), the ICM mass fraction (\fgas), the total baryon fraction (\fb), the cold baryon fraction (\fcold; hereafter we refer to this as the collapsed baryon fraction) and the metal enrichment of the ICM.  This study showed an increasing \fb\ and decreasing \fstar\ and \fcold\ in the more massive halos, suggesting that the star formation efficiency is higher in the low mass halos as well as that feedback associated with this enhanced star formation was having a larger structural impact in low mass than in high mass halos. Over the last decade, additional studies using larger samples and better data have largely confirmed this result (e.g., \citealt[][]{gonzalez07,giodini09, andreon10}; \citealt[][hereafter Zha11]{zhang11}; \citealt[][hereafter GZ13]{gonzalez13}).

Understanding how the cluster and group baryon components evolve with redshift is a key question today.  While there have been many detailed studies of intermediate and high redshift galaxy clusters, most previous observational studies of large cluster samples have focused on nearby systems due to the difficulty of defining high redshift samples and of following them up in the X-ray and with adequately deep optical or near-infrared (NIR) imaging.  That is changing now with the recent analyses of Sunyaev-Zel'dovich effect \citep[][hereafter SZE]{sunyaev70,sunyaev72} selected clusters and groups at intermediate and high redshift.  The SZE results from inverse Compton interactions of the hot ionized ICM with cosmic microwave background (CMB) photons; because it is a CMB spectral distortion rather than a source of emission, it does not suffer from cosmological dimming.  Since the first SZE selected clusters were discovered in the SPT-SZ survey \citep{staniszewski09}, this method has been demonstrated to be a useful tool for discovering and studying galaxy cluster populations out to high redshift (\citealt[][]{zenteno11}; \citealt[][hereafter H13]{hilton13}; \citealt[][]{bayliss14}).  In addition, NIR selected clusters and groups at high redshift are now also being used to study the evolution of galaxy populations \citep[e.g.][hereafter vdB14]{burg14}.  In this work we focus on an SZE selected cluster sample at redshift higher than 0.6 that originates from the first 720~$\deg^2$ of the South Pole Telescope \citep[][]{carlstrom11}  SZE (SPT-SZ) survey \citep{song12b,reichardt13}.  

To study the evolution of \fstar\ one needs robust stellar and virial mass estimates.  Stellar masses are typically estimated by converting the observed galaxy luminosity into the stellar mass using the mean mass-to-light ratio constructed from theoretical models.  This approach is sensitive to the galaxy spectral templates and needs to be modelled carefully to reduce possible biases (vdB14). For accurate stellar mass measurements with less model-dependence, one requires deep multi-wavelength observations that allow the spectral energy distribution (SED) to be measured on a galaxy by galaxy basis.  For clusters at $z\approx1$, this typically requires photometry using 8~m telescopes like the VLT together with space-based NIR data from the \Spitzer\ Space Telescope.

The cluster virial mass measurements typically have come from X-ray mass proxies such as the emission weighted mean temperature or from galaxy velocity dispersions.  The calibration of the X-ray mass proxies has often been based on the assumption of hydrostatic equilibrium, which in some circumstances can underestimate the mass by 20 -- 40\percent\ due to the non-thermal pressure components in these young structures \citep[see ][and references therein]{molnar10,chiu12}.  Velocity dispersion mass estimates, although likely less biased than hydrostatic mass estimates, have been shown to have quite high scatter on a single cluster basis \citep[e.g.,][]{white10,saro13,ruel14}.  Therefore, a study of the redshift variation of \fstar\ would benefit from a low scatter mass proxy from the X-ray or SZE that has been calibrated to mass using low bias measurements such as weak lensing or velocity dispersions together with a method that accounts for selection effects and cosmological sensitivity.  The masses we use in this analysis are based on the SZE signal-to-noise for each cluster as observed in the SPT-SZ survey and are calibrated in just such a manner \citep{bocquet14}.

In addition to robust, low scatter mass estimates one should use a uniformly selected cluster sample whose selection is not directly affected by variations in \fstar.  ICM based observables such as the X-ray luminosity or the SZE signature enable this, although connections between the physics of star formation and the structure of the ICM remain a concern.  Also, if one wishes to probe the regime beyond the group scale at high redshift, one must survey enough volume to find significant numbers of the rare, massive clusters.  Large solid angle SZE surveys like those from SPT, the Atacama Cosmology Telescope \citep[ACT,][]{fowler07} and \PLANCK\ \citep{tauber00} provide a clean way to discover clusters.  Indeed, because the SZE signature for a cluster of a given mass evolves only weakly with redshift in an arcminute resolution SZE survey, the SPT-SZ survey provides a cluster sample that is well approximated as a mass--limited sample above redshift $z\approx0.3$ \citep[e.g.][]{vanderlinde10}.  

In this paper, we seek to study the baryon content, including the ICM and the stellar mass components, of massive high redshift clusters discovered within the SPT-SZ survey. We attempt also to constrain the evolution of the baryon content of these clusters by combining our high redshift, massive clusters with other samples, primarily studied at low redshift. The paper is organized as follows. We describe the cluster sample and the data in Section~\ref{sec:samples_data}. In Section~\ref{sec:method} we provide detailed descriptions of the ICM, the stellar mass and the total mass measurements for the clusters.  We present the stellar mass function (SMF) in Section~\ref{sec:cluster_smf} and present results on the mass and redshift trends of the baryon composition in Section~\ref{sec:baryon_composition}.  We discuss these results in Section~\ref{sec:discussion} and summarize our conclusions in Section~\ref{sec:conclusion}. 

We adopt the concordance \LCDM\ cosmological model with the cosmological parameters measured in \citet{bocquet14} throughout this paper:  $\OmegaM = 0.299$, $\OmegaL = 0.701$ and $\Hnow = 68.3$~km\,s$^{-1}$\,Mpc$^{-1}$.  These constraints are derived from a combination of the SPT-SZ cluster sample, the \PLANCK\ temperature anisotropy, \emph{WMAP} polarisation anisotropy and Baryon Acoustic Oscillation (BAO) and SN Ia distances.  Unless otherwise stated all uncertainties are indicated as $1\sigma$, the quantities are estimated at the overdensity of 500 with respect to the critical density ($\rho_{\mathrm{crit}}$) at the cluster's redshift, all celestial coordinates are quoted in the epoch J2000, and all photometry is in the AB magnitude system.

%
%

\section{Cluster Samples and Data}
\label{sec:samples_data}

In this section we briefly summarize the SPT cluster sample and the follow-up data acquisition, reduction, calibration as well as the literature cluster sample we compare to. The deep optical observations from the VLT and the \HST, together with the near-infrared observations from the \Spitzer, enable us to measure the integrated stellar masses of our clusters accurately.  The ICM masses are extracted from \CHANDRA\ and \XMMNEWTON\ X-ray observations. Cluster total masses are derived from the SPT SZE observable $\xi$ as calibrated using the external data sets (see Section~\ref{sec:virial_mass}). The literature sample we compare with in this study is described in Section~\ref{sec:literature_sample}.

\subsection{SPT Cluster Sample}
\label{sec:sample}

The 14 clusters we analyze are drawn from early SPT-SZ cluster catalogs, which covered the full 2500~deg$^2$ with shallower data \citep{williamson11} or included the first 720 deg$^2$ of the full depth SPT-SZ survey \citep{reichardt13}.  The full 2500 deg$^2$ catalog has meanwhile been released \citep{bleem14}.  These 14 systems have high detection significance ($\xi > 4.8$) and were selected for further study using \HST\ and the VLT.  All fourteen have measured spectroscopic redshifts \citep{song12b}.  

\begin{figure*}
\centering
\subfigure[SPT-CL~J2331-5051 at z=0.576]{\includegraphics[width=0.4\textwidth]{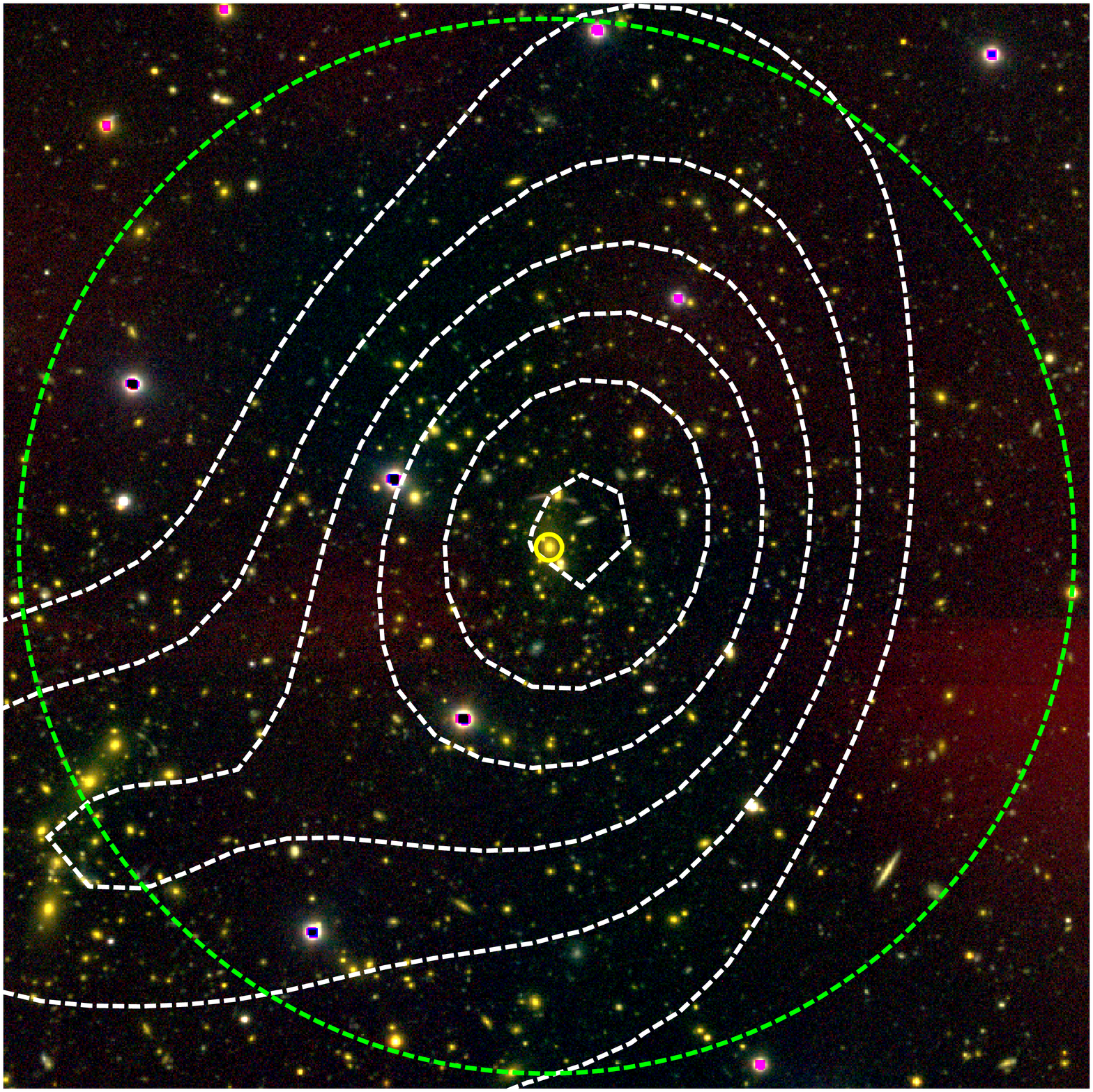}}
\subfigure[SPT-CL~J2331-5051 (zoom in)]{\includegraphics[width=0.4\textwidth]{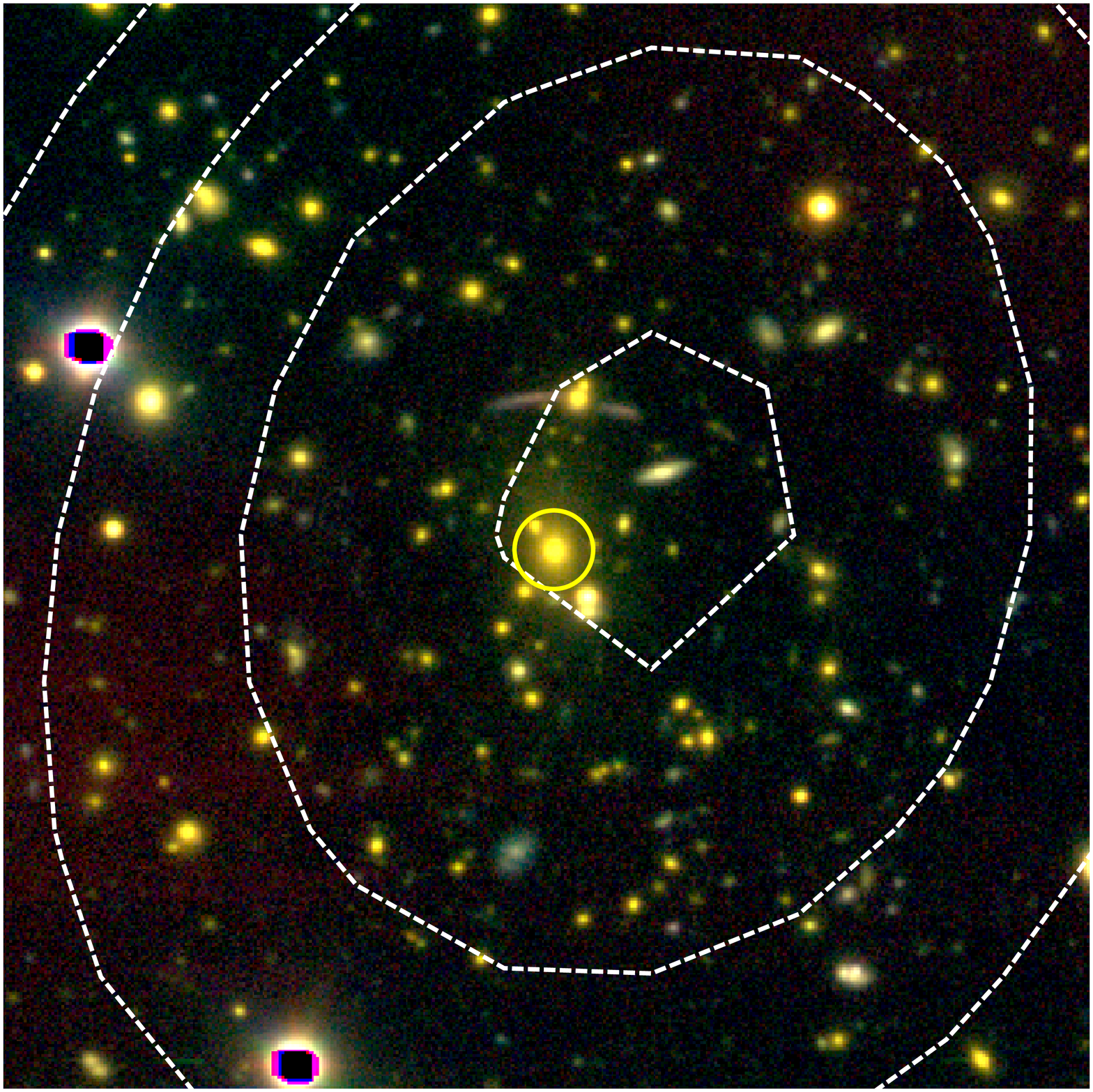}}
\caption{
VLT pseudo-color images of SPT-CL~J2331$-$5051 constructed from \BHigh, \IBess\ and \ZGunn. The left and right panels respectively show cluster \Rfiveoo\ and \Rfiveoo/3 regions centered on the X-ray peak. The SZE signal-to-noise contours from 0 to 10 with steps of 2 are white, the \Rfiveoo\ region is the green circle and the BCG is marked by the yellow circle. 
The VLT pseudo-color images for the other thirteen clusters are available online.
}
\label{fig:image}
\end{figure*}

We study the virial region defined by \Rfiveoo\ in each cluster, where \Rfiveoo\ is extracted from a virial mass estimate (\Msz) that is derived from the SPT SZE observable (see Section~\ref{sec:virial_mass}).   We adopt the X-ray centroid as the cluster center, because the SZE cluster center measurement uncertainties are larger.  A previous analysis of the offset between the SPT measured cluster center and the Brightest Cluster Galaxies (BCG) positions in a large ensemble of the SPT clusters \citep{song12b} indicated that once the SPT positional measurement uncertainties are accounted for, this offset distribution is consistent with that seen in local samples where the X-ray center is used \citep[e.g.,][]{lin04b}.  In our sample the BCG positions, X-ray centers and SZE centers are all in reasonably good agreement (see Figure~\ref{fig:image}).  Importantly, these offsets have a negligible impact on our analysis, because we are comparing average properties determined within the radius \Rfiveoo.

We present the names, redshifts and the sky positions in J2000 $(\alpha,\delta)$ of the X-ray center and BCG of our SPT sample in Table~\ref{tab:cluster_sample}.  The virial mass \Msz\ and the virial radius \Rfiveoo\ for each cluster are listed in Table~\ref{tab:results}. 

\begin{table*}
\centering
\caption{Cluster properties and photometric depths: The columns contain the cluster name, redshift and coordinates of the X-ray center and BCG
followed by the 10$\sigma$ depths in each band.}
\label{tab:cluster_sample}
\begin{tabular}{lccccccccccc}
\hline \hline 
 Cluster   & Redshift 
& $\alpha_{\mathrm{ X}} \left[\deg\right]$ 
& $\delta_{\mathrm{ X}} \left[\deg\right]$ 
& $\alpha_{\mathrm{ BCG}} \left[\deg\right]$ 
& $\delta_{\mathrm{ BCG}} \left[\deg\right]$ 
&$m^{10\sigma}_{\BHigh}$
&$m^{10\sigma}_{\Fband}$
&$m^{10\sigma}_{\IBess}$
&$m^{10\sigma}_{\ZGunn}$
&$m^{10\sigma}_{\Cha}$
&$m^{10\sigma}_{\Chb}$  \\  [3pt]
\hline  \hline
SPT-CL~J0000$-$5748 & 0.702  &  0.2500    & $-$57.8093 & 0.2502    & $-$57.8093 & 23.61 & 26.36 & 24.94 & 24.19 & 22.04 & 20.79  \\ 
SPT-CL~J0102$-$4915 & 0.870  & 15.7340   & $-$49.2656 & 15.7407    & $-$49.2720 & 24.34 & 26.31 & 24.51 & 24.14 & 22.21 & 21.86  \\ 
SPT-CL~J0205$-$5829 & 1.320  & 31.4437   & $-$58.4856 & 31.4511      & $-$58.4801   & 24.51 & 26.44 & 24.54 & 23.74 & 22.21 & 20.76  \\ 
SPT-CL~J0533$-$5005 & 0.881  & 83.4060   & $-$50.0965 & 83.4144      & $-$50.0845   & 24.64 & 26.84 & 24.66 & 23.99 & 22.01 & 20.56  \\ 
SPT-CL~J0546$-$5345 & 1.067  & 86.6548   & $-$53.7590 & 86.6569      & $-$53.7586   & 24.64 & 26.56 & 24.51 & 23.71 & 21.86 & 20.89  \\ 
SPT-CL~J0559$-$5249 & 0.609  & 89.9329   & $-$52.8266 & 89.9300      & $-$52.8242   & 24.49 & 26.46 & 24.31 & 24.06 & 22.09 & 20.71  \\ 
SPT-CL~J0615$-$5746 & 0.972  & 93.9570   & $-$57.7780 & 93.9656      & $-$57.7802   & 24.49 & 26.26 & 24.11 & 23.86 & 21.99 & 20.76  \\ 
SPT-CL~J2040$-$5726 & 0.930  & 310.0631 & $-$57.4287 & 310.0552    & $-$57.4208   & 24.51 & 26.34 & 24.69 & 24.21 & 22.24 & 20.76  \\ 
SPT-CL~J2106$-$5844 & 1.132  & 316.5179 & $-$58.7426 & 316.5192    & $-$58.7411   & 24.84 & 26.24 & 24.61 & 23.71 & 22.31 & 20.49  \\ 
SPT-CL~J2331$-$5051 & 0.576  & 352.9634 & $-$50.8649 & 352.9631    & $-$50.8650   & 24.04 & 26.41 & 24.94 & 23.51 & 22.29 & 20.71  \\ 
SPT-CL~J2337$-$5942 & 0.775  & 354.3523 & $-$59.7056 & 354.3650    & $-$59.7013   & 24.66 & 26.36 & 24.59 & 23.86 & 22.24 & 20.86  \\ 
SPT-CL~J2341$-$5119 & 1.003  & 355.3000 & $-$51.3287 & 355.3014    & $-$51.3291   & 24.59 & 26.24 & 24.81 & 23.89 & 22.26 & 20.49  \\ 
SPT-CL~J2342$-$5411 & 1.075  & 355.6916 & $-$54.1849 & 355.6913    & $-$54.1847   & 24.46 & 26.31 & 24.31 & 23.91 & 22.26 & 20.64  \\ 
SPT-CL~J2359$-$5009 & 0.775  & 359.9327 & $-$50.1697 & 359.9324    & $-$50.1722   & 24.84 & 26.19 & 24.66 & 23.91 & 21.74 & 20.66  \\ 
\hline \hline
\end{tabular}
\end{table*}

\subsubsection{Optical and Infrared Photometry}
\label{sec:optical_data}

VLT/FORS2 imaging in the bands b$_ {\mathrm{High}}$ (\BHigh), I$_{\mathrm{Bessel}}$ (\IBess), and z$_{\mathrm{Gunn}}$ (\ZGunn) was obtained for the fourteen clusters under programs 088.A-0889 and 089.A-0824 (PI Mohr). Observations were carried out in queue mode under clear conditions. The nominal exposure times for the different bands are 480\,s (\BHigh), 2100\,s (\IBess) and 3600\,s (\ZGunn). These exposure times are achieved by coadding dithered exposures with 160\,s (\BHigh), 175\,s (\IBess), and 120\,s (\ZGunn). Deviations from the nominal exposure times are present for some fields due to repeated observations when conditions violated specified constraints or observing sequences that could not be completed during the original planned semester.  The pseudo-color images of the 14 SPT clusters constructed from VLT bands \BHigh, \IBess, and \ZGunn\ are shown in Figure~\ref{fig:image}.  Each image shows also SZE contours (white), the \Rfiveoo\ virial region (green circle) and the BCG (yellow circle).

Data reduction is performed with the THELI pipeline \citep{erben05,schirmer13}.  
Twilight flats are used for flatfielding. The \IBess- and \ZGunn-band data are defringed using fringe maps extracted from night sky flats constructed from the data themselves. To avoid over-subtracting the sky background, the background subtraction is modified from the pipeline standard as described by \citet{applegate12}.

The FORS2 field-of-view is so small that only a few astrometric standards are found in the common astrometric reference catalogs. Many of them are saturated in our exposures. While we use the overlapping exposures from all passbands to map them to a common astrometric grid, the absolute astrometric calibration is adopted from mosaics of \Fband\ images centered on our clusters from the complementary ACS/\HST\ programs C18-12246 (PI Stubbs) and C19-12447 (PI High).  Each cluster is observed in the well-dithered mode through \Fband\ and F814W filters. For \Fband\ imaging, the cluster is imaged by four pointings with minimal overlap to remove the chip gap; these mosaics span a field of view of 6.7$\times$6.7 arcmin$^2$ centered on the cluster core. For F814W imaging, only one pointing centered on cluster core is acquired.  In this work we use only the \Fband\ observations for deriving the stellar masses.

Cataloging of the VLT images is carried out using SExtractor \citep{bertin96} in dual image mode.  The detection image is created through the combination of \IBess\ and \ZGunn.  Cataloging of the \HST\ images is carried out separately, also using SExtractor.  Galaxy photometry is extracted using MAG\_AUTO.  The VLT and HST photometry is matched at the catalog level with a 1\arcsec\ match radius.

Because VLT data are generally not taken in photometric conditions, the photometric calibration is also carried out using data from the \HST\ programs. We derive a relation between F814W magnitudes and the FORS2 \IBess\ filter
\begin{equation}
\label{eq:photometry_convert}
  m_{\IBess} - m_{\mathrm{F814W}} = -0.052 + 0.0095 (m_{\mathrm{F606W}} - m_{\mathrm{F814W}})  \nonumber\;,
\end{equation}
from the \cite{pickles1998} stellar library, which is valid for stars with $(m_{\mathrm{F606W}} - m_{\mathrm{F814W}}) < 1.7$\,mag.  After deriving the absolute photometric calibration of the FORS2 \IBess\ passband from this relation, the relative photometric calibrations of the other bands are fixed using a stellar locus regression \citep[e.g.][]{high09,desai12} in the $(m_{\BHigh}, m_{\Fband}, m_{\IBess}, m_{\ZGunn})$ color-space. The inclusion of \Fband\ data in this process is necessary because the stellar locus in $(m_{\BHigh}, m_{\IBess}, m_{\ZGunn})$ colors has no features.

All our clusters were observed with the \Spitzer\ Infrared Array Camera (IRAC; \citealt{fazio04}) at both 3.6~$\micron$ and 4.5~$\micron$ under programs PID 60099, 70053 and 80012 (PI Brodwin). The images are acquired in dithered mode with exposure times of $8\times100$~s and $6\times30$~s for 3.6~$\micron$ and 4.5~$\micron$, respectively.  We follow standard data reduction procedures to reduce the IRAC observations \citep{ashby09}.  For each field we generate a pair of spatially registered infrared mosaics: a relatively deep 3.6~$\micron$ image and a shallower 4.5~$\micron$ image.  These images are cataloged with SExtractor in dual image mode, using the 3.6~$\micron$ mosaic as the detection image. We use the SExtractor MAG\_AUTO and its associated uncertainty.  We verify our detections by visually inspecting the SExtractor object check image. Because the IRAC point spread function is significantly larger than in either the \HST\ or VLT imaging, we match our two-band IRAC photometry ($\Cha$ and $\Chb$) to the nearest optical counterpart at the catalog level, using a 1\arcsec\ match radius. If an object has multiple matches within the \Spitzer\ point spread function, we then deblend the IRAC fluxes into the counterparts as described below. 

For the objects in the \Spitzer/IRAC catalog with multiple optical counterparts, we deblend the \Cha\ and \Chb\ fluxes using the properties of the optical counterparts in \ZGunn.  Specifically, we deblend the \Spitzer/IRAC fluxes assuming the flux ratios of the neighboring objects in the IRAC band are the same as in the reddest optical band:
\begin{equation}
\label{eq:deblend}
R^{\Cha, \Chb}_{(i,j)} \equiv \frac{f_i}{f_j}\bigg|_{\Cha, \Chb} = \frac{f_i}{f_j}\bigg|_{\ZGunn} \, ,
\end{equation}
where $f_i$ is the flux of object $i$. 

We test the relationship between the flux ratios in \ZGunn\ and the two IRAC bands by estimating the flux ratios of matched objects without close optical neighbors.  We find that the intrinsic scatter of $R^{\Cha}_{(i,j)}$ and $R^{\Chb}_{(i,j)}$ are of the order of 0.6 and 0.8 dex, respectively. We add this scatter into the flux uncertainties in \Cha\ and \Chb\ of deblended objects. 

Although the uncertainties in the deblended fluxes are large, we find that adding these two IRAC bands-- deblended using our method-- reduces the uncertainties of the stellar mass estimates by a mean value of 20\percent\ and reduces the lognormal scatter of the reduced $\chi^{2}$ (Section~\ref{sec:sedfit}) by 29\percent.  Moreover, through studying an ensemble of pairs of  unblended sources that we first artificially blend and then deblend, we find that our method does not introduce biases in the resulting mass estimates.

The fraction of blended IRAC sources lying projected within \Rfiveoo\ for the 14 clusters varies from 11 to 20\percent\ with a mean of 16\percent\ and a standard deviation of 2.3\percent. From  25 to 55\percent\ of the sources are blended within 0.5\Rfiveoo.  Thus, the majority ($>80\percent$) of sources used in our analysis is not affected by flux blending.

We derive 10$\sigma$ depth $m^{10\sigma}_{\mathrm{filter}}$ for 6 passbands ($\mathrm{filter}=\BHigh,\Fband,\IBess,\ZGunn,\Cha,\Chb$) of each cluster in the catalog stage by estimating the magnitude where the median of the MAG\_AUTO error distribution is equal to 0.11. These values are listed in Table~\ref{tab:cluster_sample}. The $m^{10\sigma}_{\mathrm{filter}}$ depths show good consistency to the 10$\sigma$ depths estimated by measuring the sky variance in 2\arcsec\ apertures within the VLT images. The $m^{10\sigma}_{\Cha}$ depths are about 2 magnitudes deeper than our estimated \mstar\ for each cluster, and hence the cluster galaxies should be detected without significant incompleteness.

We estimate the \mstar\ of each passband for each cluster using a Composite Stellar Population (CSP) model \citep{bruzual03}.  This model has a burst at $z=3$ that decays exponentially with $e$-folding timescale of $\tau=0.4$~Gyr.  The tilt of the red sequence is modelled by using 6 CSPs with different metallicities and by calibrating those models to reproduce the Coma red sequence \citep[for more details see][]{song12a}.  This model has been shown to be adequate to derive accurate red sequence redshifts within SPT-selected clusters to $z>1$ with the root-mean-square of the cluster's photo-z error $\Delta z / (1+z)$, calibrated with spectroscopic clusters, of 0.02 \citep[][Hennig in prep.]{song12b}.  This model provides a good representation of the color and tilt of the red sequence and the evolution of \mstar\ in SPT selected galaxy clusters extending to $z\approx1.2$ (Hennig in prep.).  

\subsubsection{X-ray Data}
\label{sec:xraydata}

Eleven out of the fourteen clusters in our sample have been targeted by the \CHANDRA\ X-ray telescope with program Nos. 12800071, 12800088, and 13800883. The remaining three clusters, SPT-CL~J0205$-$5829 \cite[$z = 1.32$; see][]{stalder13}, SPT-CL~J0615$-$5746 ($z = 0.972$) and SPT-CL~J2040$-$5726 ($z = 0.93$) have been observed with \XMMNEWTON\ with program 067501 (PI Andersson).  The X-ray follow up observations are designed to observe the SPT clusters uniformly with the goal of obtaining between 1500 and 2000 source photons within \Rfiveoo.  These photons enable us to measure the ICM projected temperature, the density profile and the mass proxy $Y_{\mathrm{X}}$(the product of the ICM mass and X-ray temperature) with $\sim$15\percent\ accuracy.

The \CHANDRA\ data reduction is fully described in previous publications  \citep{andersson11, benson13,mcdonald13}.  We include an additional cluster with Chandra data (ObsID 12258), the massive merging cluster SPT-CL J0102-4915 \citep{menanteau12, jee14} at $z=0.87$, which we analyze in an identical way to those previous works (Benson et al., in prep).  For the \XMMNEWTON\ data, we use SAS 12.0.1 to reduce and reprocess the data.  All three cameras (MOS1, MOS2 and pn) are used in our analysis. The background flare periods are removed in both hard and soft bands using 3$\sigma$ clipping after point source removal. We describe the ICM mass measurements in Section~\ref{sec:icm_method}.

%
%

\subsection{Comparison Samples For This Study}
\label{sec:literature_sample}

To place our results in context and to have a more complete view of the possible redshift variation of the baryon content in galaxy clusters, we compare our measurements with the published results from the local universe at $z \le 0.1$.  We include L03, Zha11 and GZ13 because they all provide estimates of \fstar, \fgas\ and \fb\ for large samples over a broader mass range than we are able to sample with the SPT selected clusters.  L03 study 27 nearby galaxy clusters selected by optical/X-ray with masses ranging from $10^{14}$--$10^{15}\Msun$; 13 of these have available ICM mass measurements \citep{mohr99}.  There are 19 clusters in Zha11, in which \Msz\ is estimated using velocity dispersions.  We discard two clusters, A2029 and A2065, from Zha11 because they argue the virial mass estimates are biased due to the substructures.  GZ13 estimate mass fractions for 15 nearby clusters, 12 of those have stellar mass measurements. 
In addition, we include the clusters and groups from H13 and vdB14 that extend to $z \ge 0.8$, allowing a more complete study at high redshifts.  H13 study the stellar composition of 10 SZE selected clusters from ACT, and vdB14 study the Gemini CLuster Astrophysics Spectroscopic Survey (GCLASS) sample, consisting of 10 low mass clusters selected by \Spitzer/IRAC imaging.  We restrict the cluster sample to those with virial masses above $3\times10^{14}\Msun$, which is the mass regime probed by the SPT-SZ sample. This results in a total of 34 clusters in the comparison sample. We note that the majority of the vdB14 sample is in the low mass regime and therefore falls below our mass threshold; our results should not be extrapolated into this lower mass regime.

There are several important differences between these studies and ours.  We note that the groups or the clusters in these samples, with the exception of  those in H13, are either selected from X-ray or optical/NIR surveys.  Thus, these differences in selection method could potentially lead to observable differences in the samples.  In addition to these selection differences, there are differences in the stellar mass and virial mass estimates.  We describe below the corrections we apply to the comparison sample to address these differences, thereby enabling a meaningful combination with the SPT sample.

\subsubsection{Correcting to a Common IMF}
\label{sec:IMF_correction}

The most important systematic factor for estimating stellar mass is the choice of the Initial Mass Function (IMF) for the stellar population models that are then employed when converting from galaxy light to galaxy stellar mass.  Different assumed IMFs introduce systematic shifts in the mass to light ratios of the resulting stellar populations \citep{cappellari06}.  For instance, the conventional \citet{salpeter55} IMF with a power law index of -2.35 would predict a mass to light ratio higher by a factor of 2 than the one using the \cite{kroupa01} IMF \citep{kauffmann03,chabrier03,cappellari06}.  For this analysis we adopt the \citet{chabrier03} IMF (see more detailed discussion in Section~\ref{sec:stellar_method}) and apply a correction to the literature results so all measured stellar masses are appropriate for this IMF.  Specifically, we reduce the stellar mass measurements of L03 and Zha11 by 24\percent\ \citep{lin12, zhang12}, the measurements of GZ13 by 24\percent\ (or 0.12 dex), and the measurements of H13 by 42\percent\ (or 0.24 dex). Because vdB14 use the same Chabrier IMF as in this work, no IMF correction is needed. 

\subsubsection{Correcting for Virial Mass Systematics}
\label{sec:mass_systematics}

To enable a meaningful comparison of the baryon content across samples, it is crucial to use a consistent virial mass estimate for all samples.  Zha11, H13 and vdB14 estimate \Msz\ using velocity dispersions, while the other analyses all use X-ray mass proxies (ICM temperature) to estimate virial masses.  Our SPT masses arise from a recent analysis \citep{bocquet14} that includes corrections for selection effects, marginalization over cosmological parameters and systematic uncertainties and combination with external cosmological datasets (see discussion in Section~\ref{sec:virial_mass}).  

The \citet{bocquet14} analysis quantifies the systematic mass shifts that result for SPT clusters when using only X-ray data, only velocity dispersion data or the full combined dataset of X-ray, velocity dispersions and external cosmological constraints from CMB, BAO and SNe.  Namely, when compared to our cluster mass estimates obtained using the full combined dataset, the SPT cluster masses inferred from the X-ray mass proxy $Y_{\mathrm{X}}$ alone have a systematically lower mass by 44\percent,  and masses inferred from velocity dispersions alone have systematically lower masses by 23\percent.  As explained in more detail in Section~\ref{sec:virial_mass}, we adopt the full combined dataset masses for the analysis of our SPT cluster sample.  

For the comparison here, it is not crucial to know which virial mass estimate is most accurate.  What we must do is adopt one mass calibration method for our SPT sample and then correct the virial mass estimates in the other samples to a consistent mass definition before making comparisons of the baryon content.  To make these corrections we use the results from the recent SPT mass analysis \citep{bocquet14} to apply a correction to the virial mass scale in each literature sample to bring it into better consistency with our SPT sample. 

Specifically, we estimate the \Msz\ of the clusters in L03 by using the same $T_\mathrm{X}-\Msz$ relation \citep{vikhlinin09b} used in GZ13; then we increase the L03 and GZ13 masses by 44\percent, assuming the systematic offset of $Y_\mathrm{X}$ derived SPT virial masses is the same for these clusters whose masses were derived using the $T_\mathrm{X}-\Msz$ relation.  Similarly, we increase the masses in  Zha11, H13 and vdB14 by 23\percent, because their masses are derived from velocity dispersion measurements.  

Increasing \Msz\ increases the virial radius and therefore also increases the stellar and ICM masses.  Specifically, a 44\percent\ (23\percent) increase in virial mass leads to a 13.2\percent\ (7.4\percent) and 12.9\percent (7.1\percent) increment in \Mstar\ and \Mgas, respectively, assuming that the cluster galaxies are distributed as an NFW model with concentration $c_{500}=1.9$ and the ICM near the virial radius falls off as a $\beta$-model \citep{cavaliere78} with $\beta=2/3$.  In correcting the literature results for comparison to the SPT sample, we apply a correction that accounts for the shifts in all the different masses.

Correcting previously published masses to account for different data sets and analysis methods allows us to more accurately compare the results, but this correction procedure has inherent uncertainties.   It is challenging to quantify these remaining uncertainties, but for this analysis we adopt a systematic virial mass uncertainty of 15\percent\ (1$\sigma$) when constraining the redshift variation with the combined sample.  We return to this discussion in Section~\ref{sec:baryon_composition} where we present our fitting procedure in detail.  Also, in the conclusions we comment on the impact of adopting other systematic uncertainties.

We note in passing that if we had adopted the SPT masses calibrated only using the X-ray mass proxy $Y_{\mathrm{X}}$, the SPT cluster virial masses \Msz\ would decrease on average by a factor of $1/1.44$.  The new values for the SPT sample quantities \Mstar, \Mgas, \fstar, \fgas, \fcold\ and \fb\ can be approximated by applying the scale factors 0.87, 0.88, 1.26, 1.27, 0.99 and 1.27, respectively, to the measurements presented in Table~\ref{tab:results}.

%
%

\section{Mass Measurement Methods}
\label{sec:method}

In this section we describe the method for estimating the virial, the ICM and the stellar masses. 

\subsection{SPT Cluster Virial Mass \Msz\ Measurements}
\label{sec:virial_mass}

The virial masses (\Msz) that we use come from the mass calibration and cosmological analysis of \citet{bocquet14}.  They are derived using the SPT SZE observable $\xi$, the cluster redshift, and a combination of internal and external calibration data.  These data include direct mass information from 63 measured cluster velocity dispersions \citep[observed using Gemini South, the VLT, and the Magellan Baade and Clay telescopes, see][]{ruel14} and 16 $Y_{\mathrm{ X}}$ measurements \citep{andersson11,foley11,benson13}.  In addition, mass information derives from the 100 cluster candidates extracted from the first 720~$\deg^2$ of the SPT-SZ survey.  These SPT data are then jointly analyzed in combination with \PLANCK\ temperature anisotropy, \emph{WMAP}9 polarization anisotropy, BAO and SNIa constraints.  

As explained in \citet{bocquet14} (see Figure 2), adopting such strong external cosmological constraints has a dramatic impact on the cluster masses, pushing them higher to better match the masses expected within the preferred cosmological model, given the $\xi$ and redshift distribution of the cluster sample.  In contrast, the $Y_\mathrm{X}$ constraints prefer lower masses, and the velocity dispersions prefer masses in the middle.  By combining all the constraints one ends with a mass calibration that prefers higher masses than the masses one would obtain when using solely the $Y_\mathrm{X}$'s or velocity dispersions as calibrators (see also further discussion in Section~\ref{sec:mass_systematics}).  We adopt these masses that arise from a combination of internal and external calibration data for the analysis below.

Our SPT masses are corrected for Eddington bias that arises from the scatter between the mass and the selection variable $\xi$ and the steep cluster mass distribution.  The intrinsic scatter in mass at fixed $\xi$ is approximately 16\percent, and there is an additional measurement scatter that reaches $\approx 14\percent$ at $\xi=5$.   Final mass uncertainties include marginalization over all cosmological and scaling relation parameters. Thus, our masses and mass uncertainties include a combination of the systematic and statistical uncertainties.  Typical final mass uncertainties are $\sim20$\percent.  The masses are then used to calculate \Rfiveoo, which has a characteristic uncertainty of $\approx7\percent$. We refer the reader to \citet{bocquet14} for additional details.  The virial mass systematics correction for the comparison sample is described in Section~\ref{sec:mass_systematics}.

\subsection{ICM Mass Measurements}
\label{sec:icm_method}

In this work we adopt the X-ray ICM mass \Mgas\ measurements extracted within \Rfiveoo.  We determine the center of the cluster $(\alpha_{\mathrm{ X}}$,$\delta_{\mathrm{ X}})$ iteratively as the centroid of  X-ray emission in the 0.7 -- 2.0 keV energy band within a 250 -- 500~kpc annulus (see Table~\ref{tab:cluster_sample}).  The final centroid is visually verified on the smoothed X-ray emission map and is adjusted to match the center of the most circularly symmetric isophote if it deviates significantly from the peak.  The ICM density profile is estimated by fitting the X-ray surface brightness profile extracted in the energy range 0.7 -- 2.0 keV assuming spherical symmetry and centered on the derived centroid.  For \CHANDRA\ observations, we fit the modified single $\beta$-model (Equation~1 and Equation~2 in \citet{vikhlinin06}) to the X-ray surface brightness profile. The details of the X-ray analysis are given elsewhere \citep{andersson11,mcdonald13}. 

Because we cannot simultaneously constrain all the parameters in the modified single $\beta$-model for the \XMMNEWTON\ observations, we instead fit a single $\beta$-model for SPT-CL~J0205-5829 ($z = 1.32$), SPT-CL~J0615-5746 ($z = 0.972$) and SPT-CL~J2040-5726 ($z = 0.93$).  For these clusters the single $\beta$-model provides a good fit to the \XMMNEWTON\ X-ray surface brightness profile.  The best fit X-ray surface brightness profile then provides the radial distribution of the ICM, and we use the flux of the cluster within the 0.15 -- 1.0~\Rfiveoo\ annulus to determine the central density \citep[e.g.,][]{mohr99}.  We assume the metal abundance of the ICM is 0.3 solar, resulting in $n_{\mathrm{ e}} / n_{\mathrm{ p}} = 1.199$ and $\mu \equiv \rho_{\mathrm{ICM}}/(m_{\mathrm{ p}} n_{\mathrm{ e}}) = 1.16$, where the subscripts p and e denote proton and electron, respectively.  

To estimate \Mgas, we integrate the measured ICM profile to \Rfiveoo\ obtained from the SZE derived \Msz.  The uncertainty of \Mgas\ is estimated by propagating the uncertainties of the best-fit parameters. Deriving the X-ray temperature of the ICM free from the instrumental calibration bias can be challenging; however, the ICM mass and density profile is insensitive to the temperature \citep{mohr99} and to instrumental systematics \citep{schellenberger14,martino14,donahue14}.  Thus, we do not expect significant systematics in the ICM masses.

\subsection{Stellar Mass Measurements}
\label{sec:stellar_method}

In the sections below we describe the SED fitting to determine galaxy stellar masses and our method of making a statistical background correction.

\subsubsection{SED Fitting}
\label{sec:sedfit}
We use the multiband photometry to constrain the spectral energy distribution (SED) of each galaxy and to estimate its stellar mass.  
The photometry of the six bands (\BHigh\Fband\IBess\ZGunn\Cha\Chb) is used for each galaxy.
We use the Le Phare SED fitting routine \citep{arnouts99,ilbert06} together with a template library that consists of stellar templates \citep{pickles1998} and galaxy templates from CSP models \citep{bruzual03} derived using a \citet{chabrier03} IMF.  The systematics correction for the different IMF used in the comparison sample is described in Section~\ref{sec:IMF_correction}.  The stellar templates include all normal stellar spectra together with the spectra of metal-weak F- through K dwarfs and G through K giants.  The galaxy library includes templates that cover: (1) a wide range in metallicity $Z = 0.004, 0.008, 0.02$; (2) an e-folding exponentially decaying star formation rate with characteristic timescale $\tau = 0.1, 0.3, 1.0, 2.0, 3.0, 5.0, 10.0, 15.0, 30.0$~Gyr, (3) a broad redshift range between 0.0 and 3.0 with steps of 0.05, and (4) the \citet{calzetti00} extinction law  evaluated at $E(B-V) = 0.0, 0.2, 0.4, 0.6, 0.8, 1.0$.  Our galaxy library contains no templates with emission lines. 

We run the Le Phare routine with this template library on every object that lies projected within \Rfiveoo\ and is brighter than $\mstar + 2.0$ within the \ZGunn\ passband (except that we use \Cha\ for the two clusters at $z>1.1$). This ensures we are selecting the galaxy population in a consistent manner over the full redshift range.  For each galaxy, we adopt a uniform prior on the extinction law $E(B-V)$ between 0.0 and 1.0 and a weak, flat prior on the stellar mass between $10^{8}$\Msun\ and $10^{13}$\Msun.

For the SED fit we increase the MAG\_AUTO flux uncertainties for all 6 passbands by a factor of 2.  We estimate this correction factor by examining the {\it photometric repeatability} of the galaxies that appear in multiple individual VLT exposures \citep{desai12,liu14}.  With this correction the resulting magnitude uncertainties correctly describe the scatter in the repeated photometric measurements of the same galaxies.  Rescaling the uncertainties has no significant impact on the final result but increases the uncertainty of the stellar mass estimate for each galaxy by 25\percent.

For each cluster we first estimate the stellar mass of the BCG, \MBCG, fixing the redshift to the cluster redshift.  The BCG is chosen to be the brightest cluster galaxy projected within \Rfiveoo; we select this galaxy visually using the NIR and optical imaging and then confirm in the catalog (\ZGunn\ and \Cha) that it is the brightest galaxy.  
We find that the BCGs in our cluster sample all prefer the templates with the characteristic e-folding timescale for the star formation rate to be $\tau\le1$~Gyr.  
This indicates that the rapid star forming activity seen in the SPT selected Phoenix Cluster BCG \citep[][]{mcdonald14} is not present in our cluster sample.  This result is consistent with the view that the evolution of the typical BCG is well described by a CSP model with $\tau\approx0.9$~Gyr out to redshift 1.5 \citep{lidman12}. For the final \MBCG\ estimates we restrict the template library to $\tau\le1$~Gyr, which results in a $\approx6\percent$ reduction in the stellar mass uncertainties as compared to fitting across the full range of $\tau$.  This small change in uncertainty has no impact on our final result.
We then estimate the stellar mass for the remaining galaxies using the same configuration except that we allow the redshift to float and fit the templates without restricting $\tau$. 

We adjust the Le Phare routine to output the best-fit mass \Mbest, the median mass \Mmed, the mass at the lower (higher) 68\percent\ confidence level \Mlo\ (\Mhi) and the best-fit $\chi^{2}$ extracted over the full template library.  We discard the objects from the analysis where the best fit $\chi^{2}$ arises for a stellar template.  This stellar removal works well; testing on the COSMOS field \citep{capak2007,sander2007,ilbert2008} indicates we have a residual stellar contamination and a false identification rate for galaxies under 1.5\percent\ and $\sim$0.15\percent, respectively.  
The mass-to-light ratios $\Upsilon$ and their rms variations in the observed frame \Cha\ band for all clusters are provided in Table~\ref{tab:results}.  These are extracted from the subset of galaxies projected within the virial region that have photo-z's that are within $\Delta z=0.1$ of the cluster spectroscopic redshift.

We examine those galaxies with $\Mbest>\MBCG$ closely, because we expect no galaxy to be more massive than the BCG. We find that most of these galaxies can be excluded because they have redshifts far higher than the cluster. In total, there are 37 out of 2640 galaxies with $\Mbest > \MBCG$ within \Rfiveoo\ of the 14 clusters. That is, about 1.5\percent\ of the objects are discarded through this process. However, one must take special care in cases of merging clusters, which could host one or more galaxies with masses similar to the most massive one.  In a few cases (3 galaxies to be exact) we find through photo-z and visual inspection that these galaxies likely lie in the cluster and have mass estimates slightly larger than the mass of our selected BCG.  In these cases we include those galaxies in the stellar mass estimate. We provide further details of our SED fit performance in Appendix~\ref{sec:specz}.

The stellar mass estimate for the region within \Rfiveoo, including the foreground and background galaxies, is the sum of \Mbest.  The uncertainty includes the uncertainties on the single galaxy masses (using \Mlo\ and \Mhi\ and assuming the probability distribution for the stellar mass is Gaussian).

We also calculate the fraction of objects $f_{\mathrm{cor}}$ with unphysical mass outputs (i.e., $\log(\Mmed) = -99.0$), which occur mostly due to data corruption.  We correct for these missing galaxies by assuming that they share the same distribution of stellar masses as the uncorrupted galaxies. We note that this fraction varies between 3 and 10\percent. A correction for the masking of the bright stars is also applied.  Thus, for each cluster we estimate the total stellar mass $M_{\star}^{\mathrm{field}}$ projected within \Rfiveoo\ as 
\begin{equation}
M_{\star}^{\mathrm{field}}={\Sigma_i M_{\star,i}^{\mathrm{best}} \over 
\left (1-f_{\mathrm{mask}}\right) \left(1-f_{\mathrm{cor}}\right)} \, ,
\label{eq:mfield}
\end{equation}
where $f_{\mathrm{mask}}$ is the fraction of area within \Rfiveoo\ that is masked and $M_{\star,i}^{\mathrm{best}}$ is the best stellar mass estimate for galaxy $i$ in the cluster.

\subsubsection{Background Correction}
\label{sec:BKGsubtraction}

We correct the stellar mass from the cluster field $\Mstar^{\mathrm{field}}$ for the contribution from the foreground and background galaxies $\Mstar^{\mathrm{bkg}}$ using a statistical correction.  Because the FORS2 field of view is small, the background regions outside \Rfiveoo\ are contaminated by cluster galaxies.  Thus, we use the COSMOS survey to estimate the background correction.

The COSMOS survey has 30-band photometry with wavelength coverage from UV to mid-infrared.  To minimize systematics we take two steps to make the COSMOS dataset as similar as possible to our SPT dataset.  First, we choose the passbands which are most similar to ours (from Subaru Suprime-Cam and \Spitzer) and apply color corrections where needed to convert the COSMOS photometry into our passbands. MAG\_AUTO photometry is used in the COSMOS field.  Second, we degrade the COSMOS photometry to have the same measurement noise as in our dataset.

We then measure the stellar mass for each galaxy in the COSMOS field using the converted photometry, the same \Spitzer\ object detection, the same matching algorithm, and the same fitting strategy as we applied to our own data.  We correct this background estimate for the fraction of corrupted galaxies as described for the cluster fields in Equation~\ref{eq:mfield}.  

Then, correcting the COSMOS background estimates to the area of each cluster field, we subtract the background estimate $\Mstar^{\mathrm{bkg}}$, obtaining our estimate of the cluster stellar mass projected within \Rfiveoo.  We then apply a geometric factor $f_{\mathrm{geo}}$ to correct this projected quantity to the stellar mass within the virial volume \Mstar\ using a typical radial galaxy profile with concentration $c^{\mathrm{gal}}_{500}=1.9$ \citep[][Hennig in prep]{lin04a}, which corresponds to a normalization correction of $f_{\mathrm{geo}}=0.71$.
\begin{equation}
M_*=\MBCG+f_{\mathrm{geo}}\left(M_*^{\mathrm{field}}-M_*^{\mathrm{bkg}}\right) \, ,
\label{eq:mstar}
\end{equation}
where we have defined \Mstar\ to include \MBCG, the BCG stellar mass.

In Appendix~\ref{sec:background_test}, we compare the COSMOS background to the background estimated in the non-cluster portions of the VLT imaging where a correction for cluster contamination has been applied.  We find that the two backgrounds agree at the 10\percent\ level, leading to background corrected cluster stellar mass estimates \Mstar\ that are consistent at the 4\percent\ level.  Thus, we adopt this difference as the amplitude of the systematic uncertainty associated with our statistical background correction. 

%
%

\section{BCG and Cluster Stellar Mass function}
\label{sec:cluster_smf}

In this section we present the components of the cluster stellar mass function (SMF).  These include the BCGs (Section~\ref{sec:bcg_mass}), which we discuss first, followed by the full SMF and the luminosity function (LF) of the satellite galaxies (Section~\ref{sec:smf}).  

\subsection{BCG Stellar Mass}
\label{sec:bcg_mass}

We present \MBCG\ for the 14 SPT clusters and compare them with the measurements of H13 and vdB14, in which groups and clusters at $z\ge0.3$ are studied.  H13 estimated \MBCG\ based on the mass-to-light technique assuming a passive evolution model with the \Cha\ magnitude MAG\_AUTO. vdB14 applied the same technique using the $K_s$ luminosity together with the Sersic model fitting to the light profile.  As noted in vdB14, the magnitude inferred by the Sersic profile could differ from MAG\_AUTO by up to 0.2~mag, depending on the shape of the light profile.  In this work we estimate \MBCG\ using an SED fit to the six bands available in our survey. No special attempt is made to include or deblend the extended halo or intracluster light (ICL) in any of these studies.  We have three clusters in common with H13: SPT-CL~J0102$-$4915, SPT-CL~J0546$-$5345 and SPT-CL~J0559$-$5249.  For SPT-CL~J0546$-$5345 our \MBCG\ is about a factor of 2 higher, but in the other two clusters \MBCG\ agrees at better than 10\percent. We compare the \Cha\ photometry of SPT-CL~J0546$-$5345, and find that the magnitude reported by H13 is about 0.5 mag fainter, suggesting that a more sophisticated deblending algorithm is needed for the crowded core of SPT-CL~J0546$-$5345 in this work.  We adopt the SED mass estimates for the BCGs in the analyses that follow.
The BCGs are marked by yellow circles in the optical images presented in Figure~\ref{fig:image}.

\begin{figure}
\vskip-0.25cm
\centering
\includegraphics[scale=0.6]{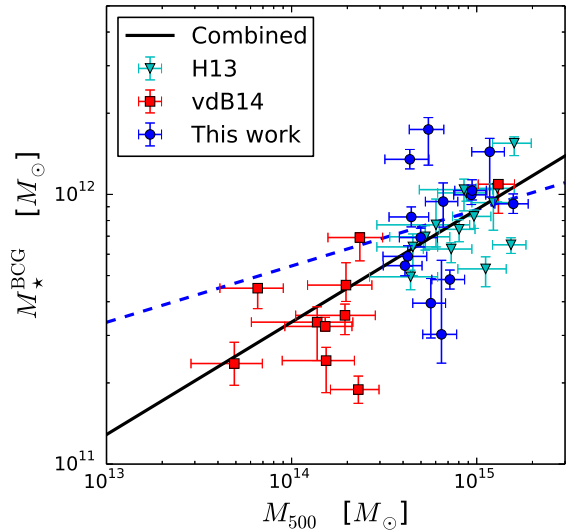}
\caption{The BCG stellar mass (\MBCG) versus cluster virial mass \Msz\ for the SPT sample (blue),  H13 (cyan) and vdB14 (red).  The H13 sample is corrected to Chabrier IMF.  The blue dashed line is the best-fit for the SPT sample alone and the black solid line is the best fit for the combined sample (see Equation~\ref{eq:bcg}).}
\label{fig:bcg_plot}
\vskip-0.5cm
\end{figure}

As is clear in Figure~\ref{fig:bcg_plot}, neither the SPT nor the H13 sample provides strong evidence for a correlation between the BCG mass and the cluster halo mass.  However, in combination with the vdB14 sample that extends to much lower mass, we find a best fit  \MBCG--\Msz\ relation of 
\begin{equation}
\label{eq:bcg}
\MBCG  =  (5.30  \pm  0.39) \times 10^{11} \left( \frac{ \Msz }{ 3 \times 10^{14} \Msun } \right)^{0.42  \pm  0.07}  \, ,
\end{equation}
for the combined sample, and this relation is plotted in Figure~\ref{fig:bcg_plot} (black dashed line).
Thus, the BCG stellar mass constitutes about 0.12\percent\ of the cluster halo mass at $\Msz=6\times10^{14}\Msun$.  Because \MBCG\ scales with cluster halo mass with a power law index less than one,  the fraction of the cluster mass made up by the BCG falls as $\MBCG/\Msz\propto \Msz^{-0.58\pm0.07}$.

The SPT sample scatters significantly about this relation, providing evidence of intrinsic scatter in \MBCG\ at fixed cluster halo mass of $\sigma_{\mathrm{int}} = 0.17 \pm 0.034$~dex. The full sample exhibits a consistent value $\sigma_{\mathrm{int}} = 0.15 \pm 0.021$~dex.  Thus, the characteristic scatter of the BCG masses at a fixed cluster halo mass is 41\percent.

\subsection{Cluster Luminosity and Stellar Mass Functions}
\label{sec:smf}

We extract the \Cha\ LF and the SMF using a statistical background subtraction with the COSMOS field as the source of the background (see Section~\ref{sec:BKGsubtraction}). We apply a correction from the virial cylinder to the virial volume in the same manner as in Section~\ref{sec:BKGsubtraction}. The measured LF and SMF are in physical density units of Mpc$^{-3}$.  The uncertainty of each bin is estimated by the Poisson error associated with the galaxy counts in the case of the LF and this error combined with the galaxy stellar mass measurement uncertainties for the SMF.

We stack the LF and SMF from 14 SPT clusters using inverse-variance weighting within each bin. The number densities are corrected to the median redshift of the SPT clusters, $z=0.9$, by multiplying by the ratio of the critical densities, $\left(\frac{E(0.9)}{E(z)}\right)^2$, where $E(z)^2\equiv \OmegaL + \OmegaM(1+z)^3$ and $z$ is the redshift of the cluster.  We stack the LF within the space of $m-\mstar$ with magnitude bins of width 0.5, where $\mstar$ comes from the CSP model described in Section~\ref{sec:optical_data}.  
Given that the galaxy population in SPT selected clusters has been shown to be well described by the CSP model (\cite{song12b}, Hennig et al, in preparation) stacking LFs with respect to the $\mstar$ predicted at the redshift of each cluster provides a simple way to extract the information for the normalization and shape of the composite LF.
We stack the SMF in the stellar mass range from $10^{10}$ -- $10^{12}\Msun$ with bin width of 0.2~dex. Finally, we characterize the stacked LF and SMF with the standard Schechter function \citep{schechter76}. Specifically, we fit the stacked LF directly in log space to:
\begin{eqnarray}
\label{eq:lf}
\Phi_L(m)                 &= &0.4 \ln(10.0) \phi^* \times 10.0^{-0.4(\alpha_{\mathrm{ L}}+1)(m-\mzero)} \nonumber \\ 
 & &\times \exp( -10.0^{-0.4(m-\mzero)} )  \, ,
 \end{eqnarray}
 where $m$ is the magnitude, \mzero\ is the characteristic magnitude predicted by the passively evolving model (see Section~\ref{sec:optical_data}), $\phi^*$ is the characteristic density and $\alpha_{\mathrm{ L}}$ is the faint end slope.  We fit the stacked SMF directly in log space to:
 \begin{eqnarray}
\Phi_{M} (M_{\star})  &= &\ln(10) \phi_M \times 10^{(\alpha_{\mathrm{ M}}+1)(m_{\star}-\Mzero)} \nonumber \\ 
 & &\times \exp( -10^{(m_{\star}-\Mzero)} ) \, ,
\label{eq:smf}
\end{eqnarray}
where $m_{\star}$ is the stellar mass in units of $\log_{10}(m_{\star}/\Msun)$, \Mzero\ is the characteristic mass, $\phi_{\mathrm{ M}}$ is the characteristic density, and $\alpha_{\mathrm{ M}}$ is the faint end slope. We restrict our fit to those galaxies brighter than $\mstar+2$ in the LF analysis. Because the stellar mass is not a linearly-rescaled version of the magnitude, we choose the conservative depth limit used in the SMF analysis, which is based on the mass-to-light-inferred mass at brighter magnitude, $\mstar+1.5$, assuming the passively evolving model for SMF analysis. 

\begin{figure}
\includegraphics[scale=0.6]{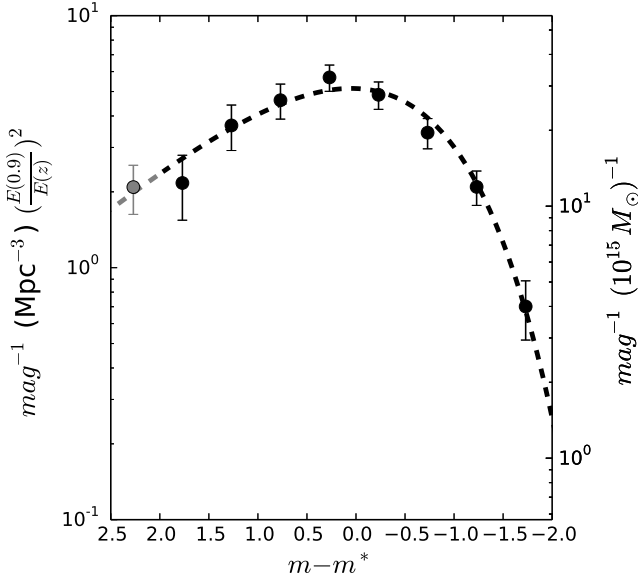} 
\vskip-0.25cm
\caption{The stacked luminosity function of 14 SPT clusters extracted from the \Cha\ photometry (black points).  The grey point is fainter than $\mstar+2$ and is not included in the fit. The line marks the best fit Schechter function.  The LF is plotted versus $m-\mstar$, where \mstar\ is obtained from the passive evolution model described in the text (Section~\ref{sec:optical_data}).  The stacked number densities are corrected for evolution of the critical density ($\rho_{\mathrm{crit}}\propto E(z)^2$) and normalized to median redshift $z=0.9$.}
\label{fig:LF}
\vskip-0.35cm
\end{figure}
\begin{figure}
\vskip-0.25cm
\includegraphics[scale=0.6]{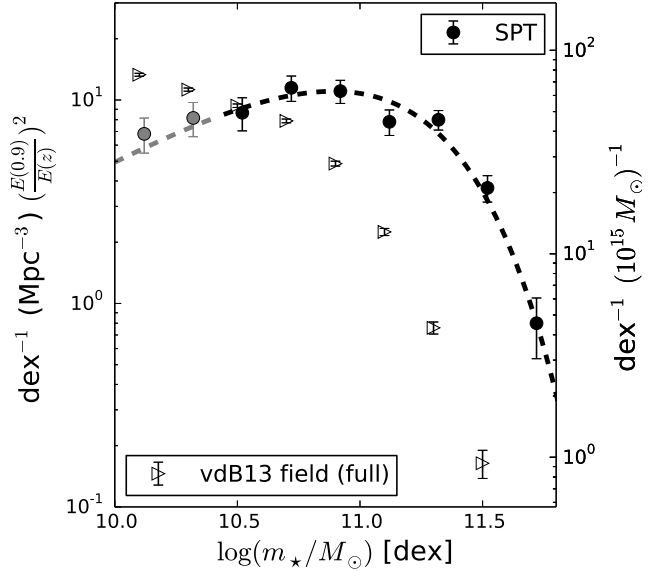}
\vskip-0.25cm
\caption{The measured stellar mass from (SMF) obtained by stacking 14 SPT clusters. The black line is thebest fit Schechter function (see Table~\ref{tab:smf}). The grey points are measurements beyond the depth limit and are not used in the fit.  For comparison, we show the field SMF from vdB13.}
\label{fig:smf}
\vskip-0.5cm
\end{figure}

The stacked  LF and SMF are shown in Figure~\ref{fig:LF} and Figure~\ref{fig:smf}, respectively. The best-fit parameters are given in Table~\ref{tab:smf}.  We convert the SMF and LF from physical number density to the abundance per mass of $10^{15}\Msun$ (total baryon and dark matter mass) by using the mean density within the virial region at $z=0.9$, which is $500\rho_{\mathrm{crit}}(z=0.9)$.  This value is shown on the right $y$-axis.  Similarly, to compare to a field LF or SMF one would convert from Mpc$^{-3}$ to per unit mass by using the mean density of the universe at that redshift $\left<\rho\right>(z)=\OmegaM(z)\times\rho_{\mathrm{crit}}(z)$.

\begin{table}
\centering
\caption{Luminosity and Stellar Mass Function Parameters: The luminosity function (top) characteristic density, characteristic magnitude, faint end slope and reduced $\chi^2$ are shown followed (below) by the equivalent stellar mass function parameters.}
\label{tab:smf}
\begin{tabular}{cccc}
\hline \hline
$\phi^*$                              & $\mzero$   &                     &                                     \\ 
$[$Mpc$^{-3}$mag$^{-1}]$ & $[$mag$]$  &  $\alpha_L$  & $\chi_{\mathrm{red}}^2$ \\ 
\hline
$14.90  \pm  1.0  $   & $-0.18  \pm  0.1  $   & $-0.19  \pm  0.1  $   & $0.6 $ \\ \hline
$\phi_M$                             & $\Mzero$   &                      &                                     \\
$[$Mpc$^{-3}$dex$^{-1}]$  & $[$dex$]$   & $\alpha_M$  & $\chi_{\mathrm{red}}^2$ \\ 
\hline
$12.30  \pm  1.5  $   & $11.06  \pm  0.1  $   & $-0.32  \pm  0.2  $   & $1.4 $ \\
\hline \hline
\end{tabular}
\end{table}

The best-fit \mzero\ indicates that the LF deviates from the predicted characteristic $\mstar_{\Cha}$ for the passive evolution model (Section~\ref{sec:optical_data}) by $-0.18\pm0.1$, suggesting the mild evidence (about 1.8$\sigma$) of the blue population at the high redshift clusters. The best-fit SMF and LF are consistent with one another; the characteristic $\mstar_{\Cha}$ at median redshift $z=0.9$ predicted by the passively evolving model corresponds to the stellar mass of $10^{10.96}\Msun$, while the measured characteristic mass is $10^{11.0\pm0.1}\Msun$.  The faint end slopes and characteristic densities are also in good agreement. 

In a recent paper, \citet[][hereafter vdB13]{burg13} compare the SMFs of the GCLASS low mass clusters to the field at redshift $z=0.85-1.2$ and find the number density of galaxies per unit mass (dark matter plus baryons) in the field SMF is lower than that in the groups over the mass range $10^{10}\Msun$ to $10^{11.5}\Msun$.  This suggests that the galaxy formation rate has been lower over time in the field than in the dense group and cluster environments.  A similar picture had previously emerged in the local Universe ($z<0.1$) \citep{lin04a,lin06}, where the luminosity functions of K-band selected galaxies and of radio sources within clusters are also significantly higher than the field after corrections for the mean matter density differences in the two environments.  As seen in Figure~\ref{fig:smf}, the normalization of the SMF for the SPT clusters on the massive end ($\log_{10}(m_{\star}/\Msun)\approx11.2-11.5$) is significantly higher than the field (open triangle) measured by vdB13. By integrating the best-fit SMF of SPT above our single galaxy stellar mass threshold of $2.5\times10^{10}\Msun$, we estimate the number of galaxies per unit total mass for SPT clusters is $\approx1.65\pm0.20$ times higher than the field at $z=0.85-1.2$.  Our result reinforces this picture that the cluster environment contains a more biased galaxy population than the field.

%
%

\section{Baryon Composition}
\label{sec:baryon_composition}

In this section we present our measurements for the stellar mass fraction, ICM mass fraction, collapsed baryon fraction and baryon fraction:
\begin{eqnarray}
\label{eq:fraction_def}
\fstar	 &\equiv	&\frac{\Mstar}{\Msz}         \\
\fgas    &\equiv	&\frac{\Mgas}{\Msz}          \\
\fcold	 &\equiv	&\frac{\Mstar}{\Mbaryon}  \\
\fb       &\equiv	&\frac{\Mbaryon}{\Msz}    \ ,
\end{eqnarray}
where $\Mstar$ is the stellar mass (see Equation~\ref{eq:mstar}), $\Mgas$ is the ICM mass (see Section~\ref{sec:icm_method}) and $\Mbaryon\ \equiv \Mstar\ + \Mgas$ is the total mass in baryons.  $\Msz$ is the halo virial mass, estimated using the SZE observable (see Section~\ref{sec:virial_mass}).

In addition, we study mass and redshift trends in our SPT clusters and in the combined sample that includes the clusters studied in the literature (discussed in Section~\ref{sec:literature_sample}). Note that we are not probing the evolution of the baryon content by directly tracing the progenitors, because our SPT sample lacks low mass groups at all redshifts. We instead estimate the baryon content of the massive clusters with respect to the characteristic mass at the different epochs statistically by fitting the scaling relation simultaneously in mass and redshift space (see Section~\ref{sec:fitting_procedure}).  We also compare our cluster results with more general results coming from external, non-cluster datasets.   We use the universal baryon fraction \fb\ estimated using the \PLANCK\ CMB anisotropy observations \citep{planck13-16}, and we estimate the universal stellar density parameter $\Omega_{\star}$, where the mean stellar density at $z=0$ is extracted from the local $K$-band galaxy LF \citep{kochanek01} and the mean stellar density at $z=1$ is extracted from the vdB13 analysis.  These values have been corrected to our fiducial cosmology and are appropriate for a Chabrier IMF, enabling comparison to our cluster measurements.

\subsection{Fitting Procedure}
\label{sec:fitting_procedure}

We fit these measurements from our cluster ensemble and also from the combined sample to a power law relation in both mass and redshift:
\begin{equation}
\label{eq:fitting_form}
f_{\mathrm{obs}}(\Msz,z) = \alpha_{\mathrm{obs}} 
\left(\frac{\Msz}{ M_{\mathrm{piv}} }\right)^{\beta_{\mathrm{obs}}}
\left(\frac{1 + z}{1+z_{\mathrm{piv}}}\right)^{\gamma_{\mathrm{obs}}}
\end{equation}
where $M_{\mathrm{piv}}$ and $z_{\mathrm{piv}}$ are the mass and redshift pivot points, $\mathrm{obs}$ corresponds to the different observables and $\alpha_{\mathrm{obs}}$, $\beta_{\mathrm{obs}}$ and $\gamma_{\mathrm{obs}}$ correspond to normalization of the best fit relation, the power law index of the mass dependence and the power law index of the redshift dependence, respectively.  We perform $\chi^2$ fitting directly in log space using the measurement uncertainties and accounting for intrinsic scatter.  For the SPT and combined samples we choose the pivot points to be the median mass $\Msz=6\times10^{14}\Msun$.  For the SPT sample we adopt the redshift pivot $z_{\mathrm{piv}}=0.9$, consistent with the median redshift of the sample, but for the combined sample we adopt a redshift pivot of $z_{\mathrm{piv}}=0$.

\begin{table}
\centering 
\caption{Mass and Redshift Trends of Baryon Composition with $M_{\mathrm{piv}} \equiv 6\times 10^{14} \Msun$:  The columns contain the quantity of interested, the normalization at the pivot mass and redshift, mass dependence and redshift dependence (see Equation~\ref{eq:fitting_form}) for the SPT sample alone (above) and for the SPT sample together with the literature sample (below).}
\label{tab:power_law}
\begin{tabular}{lrrr}
 \hline \hline
   $f_{\mathrm{obs}}$  &  $\alpha_{\mathrm{obs}}$  & $\beta_{\mathrm{obs}}$  & $\gamma_{\mathrm{obs}}$ $\dag$ \\ [3pt] \hline \hline 
\multispan{4}{\centering{ \hfil SPT Sample Results with $z_{\mathrm{piv}}\equiv0.9$ \hfil } } \\  \hline
\fstar\   & $0.011\pm0.001$  & $-0.09 \pm  0.27$   & $1.07  \pm  1.08$ \\
\fgas\    & $0.096\pm0.005$ & $0.43  \pm  0.13$   & $0.20  \pm  0.49$ \\
\fcold\   & $0.107\pm0.011$ & $-0.55 \pm  0.22$ & $0.81 \pm  0.93$  \\
\fb\       & $0.107\pm0.006$  & $0.39  \pm  0.13$   & $0.32  \pm  0.50$ \\ \hline
\multispan{4}{\centering{ \hfil Combined Sample Results with $z_{\mathrm{piv}}\equiv0$ \hfil} } \\ \hline
\fstar\  &  $0.0099  \pm  0.0005$  &  $-0.37  \pm  0.09$  &  $0.26  \pm  0.16 \pm 0.08$  \\  
\fgas\  &  $0.1120  \pm  0.0032$  &  $0.22  \pm  0.06$  &  $-0.20  \pm  0.11 \pm 0.22$  \\    
\fcold\  &  $0.0859  \pm  0.0049$  &  $-0.65  \pm  0.10$  &  $0.39  \pm  0.15 \pm 0.16$  \\    
\fb\  &  $0.1227  \pm  0.0035$  &  $0.22  \pm  0.06$  &  $-0.17  \pm  0.11 \pm 0.22$  \\  
\hline
\end{tabular}
$\dag$ \footnotesize The second $\gamma_{\mathrm{obs}}$ uncertainty arises from the 15\percent\ \Msz\ systematic uncertainty.
\end{table}

The parameters for the best-fit relations for the SPT sample and for the combined sample are listed in Table~\ref{tab:power_law}, while the measured cluster virial masses, ICM masses, stellar masses and the derived quantities above are listed in Table~\ref{tab:results}.  These results are summarised in Figures~\ref{fig:mass_trends} and \ref{fig:redshift_trends}, where the first figure focuses on the mass trends and the second focuses on the redshift trends. In the subsections below we discuss each derived quantity in turn.

\subsubsection{Accounting for \Msz\ Systematic Uncertainties}
\label{sec:fit_mass_systematic}

We account for systematic differences in \Msz\ estimation between the low redshift comparison sample (L03, Zha11 and GZ13) and the high redshift sample (SPT with two additional samples of H13 and vdB14 added when comparing \fstar) by adopting a 15\percent\ (1$\sigma$) systematic virial mass uncertainty (see discussion in Section~\ref{sec:literature_sample}).  These virial mass uncertainties imply corresponding \Rfiveoo\ uncertainties that lead also to systematic uncertainties in the stellar mass and ICM mass for each cluster.  We estimate the systematic uncertainties in the redshift variation parameter $\gamma_{\mathrm{obs}}$ (Table~\ref{tab:power_law}) by perturbing the virial masses of the high redshift sample by $\pm15\percent$ and extracting the best fit parameters in each case.  The 1$\sigma$ systematic uncertainty is estimated as half the difference between the two sets of parameters. This virial mass systematic is only important for the measured redshift trends.

\subsubsection{Accounting for Differences in Measurement Uncertainties}
\label{sec:fit_uncertainties}

We also account for systematic differences in the measurement uncertainties among the different samples by solving for a best fit intrinsic scatter separately for each sample.  For the SPT sample, where mass uncertainties include both measurement and systematic uncertainties (Section~\ref{sec:virial_mass}), we find no need for an additional intrinsic scatter.  The best fit estimates of the intrinsic scatter for the other samples are 9\percent\ for \fgas\ and \fb\ in L03, 14\percent\ for \fstar, \fgas\ and \fb\ in GZ13, 18\percent\ for \fstar\ in H13 and 20\percent\ in vdB14, and 20 to 22\percent\ for the fractions in Zha11. Three of the samples with the largest intrinsic scatter (Zha11, H13 and vdB14) employ velocity dispersions for single cluster mass estimation as opposed to X-ray or SZE mass indicators. This is not surprising, because it has been shown that cluster velocity dispersions provide high scatter single cluster mass estimates \citep[see][and references therein]{saro13}.  Velocity dispersions can be effectively used in ensemble to calibrate ICM based single cluster mass estimates \citep{bocquet14}.  

\subsection{Stellar Mass Fraction \fstar}
\label{sec:star_literature}

\begin{figure}
\centering
\includegraphics[scale=0.55]{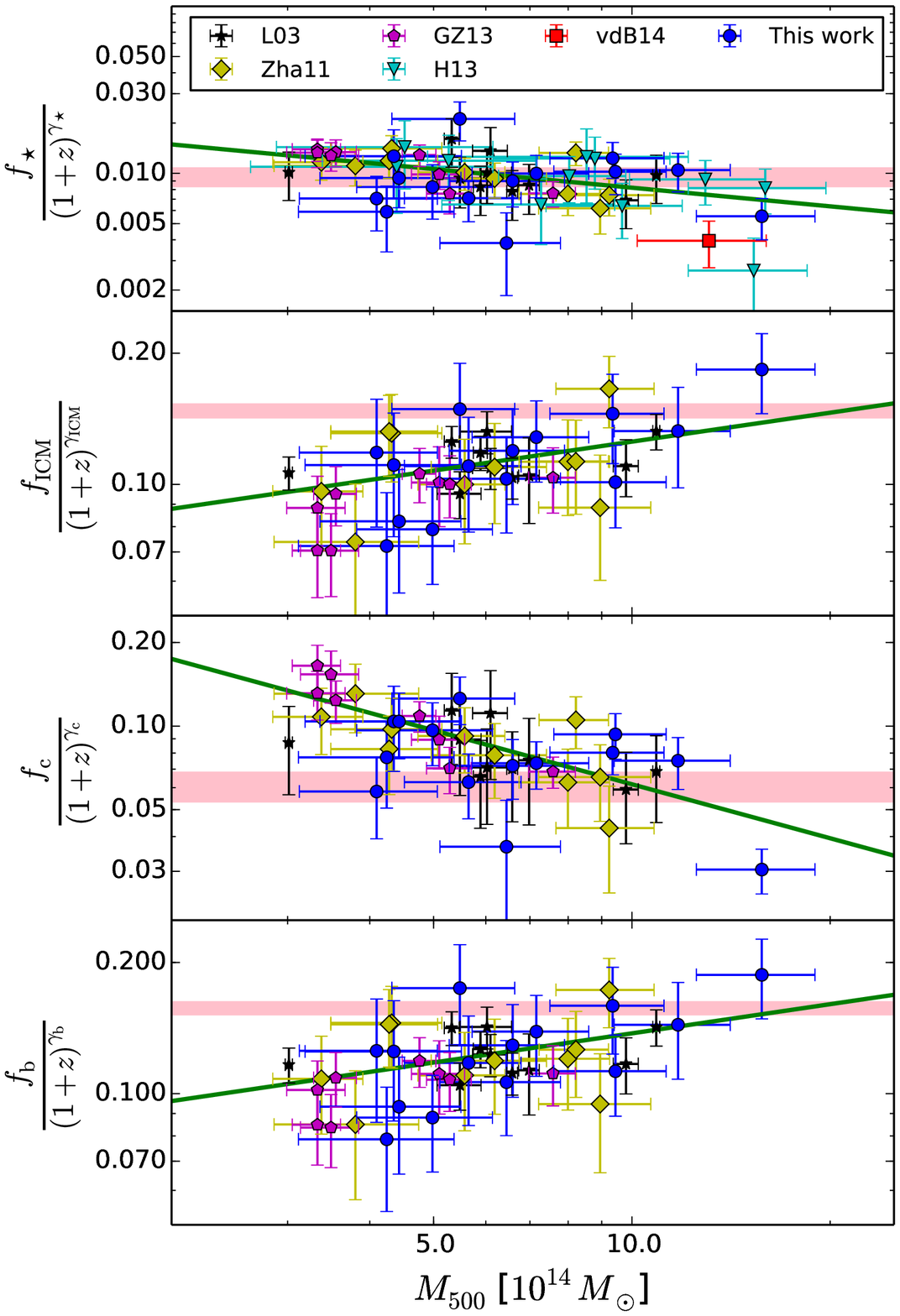}
\caption{The baryonic fractions \fstar, \fgas, \fcold\ and \fb\ are shown as a function of cluster virial mass \Msz\ for the combined sample.  In all cases the measurements have been corrected to $z=0$ using the best fit redshift trend.  The best fit mass trend is shown in green (Table~\ref{tab:power_law}). The color coding and point styles are defined in the upper panel and is the same throughout.  The red shaded region indicates the universal baryon composition from combining the best-fit cosmological parameters from \citet{bocquet14} together with the local $K$-band luminosity function \citep{kochanek01}.}
\label{fig:mass_trends}
\vskip-0.5cm
\end{figure}

The stellar mass fraction we estimate here is the mass in stars within cluster galaxies.  We make no attempt to account for the ICL component.  Figure~\ref{fig:mass_trends} contains a plot of our results (blue).  The mean \fstar\ of our fourteen clusters is $0.011 \pm 0.001$, and the characteristic value at $z=0.9$ and $\Msz=6\times10^{14}\Msun$ is $0.011  \pm  0.001$.  The SPT sample provides no evidence for a mass or redshift trend, but the large mass trend uncertainty ($\fstar\propto\Msz^{-0.09\pm0.27}$) means the sample is statistically consistent with the trend for more massive clusters to have lower $\fstar$ (L03).  In the combined sample, there is 3.7$\sigma$ evidence for a mass trend $\fstar\propto\Msz^{-0.37\pm0.09}$, which is also consistent with the L03 result.  The combined sample exhibits no significant redshift variation ($\fstar\propto(1+z)^{0.26\pm0.16\pm0.08}$),
where the second uncertainty reflects the 15\percent\ (1$\sigma$) systematic virial mass uncertainty.  The characteristic value at $z=0$ is $\fstar=0.010\pm0.0005$ (statistical), which is in good agreement with the SPT value at $z_{\mathrm{piv}}=0.9$.

\begin{figure}
\centering
\includegraphics[scale=0.55]{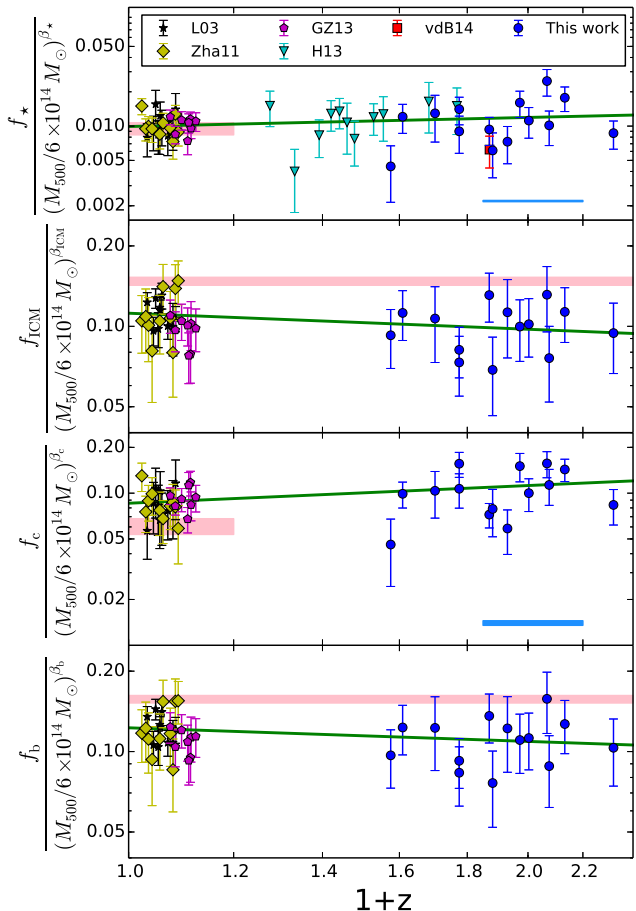}
\caption{The redshift trends of \fstar, \fgas, \fcold\ and \fb\ for the combined cluster sample.  The color coding of the points and the red bands are the same as in Figure~\ref{fig:mass_trends}.  For \fstar\ and \fcold\ we show the universal value at $z=0$ (red) and at $z=0.85-1.2$ (blue).  Measurements have been corrected using the best fit mass trends to the mass $6\times10^{14}\Msun$, and the best fit redshift trend is shown in green (Table~\ref{tab:power_law}).}
\label{fig:redshift_trends}
\vskip-0.5cm
\end{figure}

Also shown in the shaded region is the \fstar\ constraint emerging from a combination of the stellar mass density from the $K$-band local luminosity function \citep{kochanek01}, $\Omega_{\star}h = 3.4 \pm 0.4 \times 10^{-3}$ with $h = 0.683$, with the most recent combined results (\PLANCK\ + \emph{WMAP} polarization+SNe+BAO+SPT clusters) on the cosmological matter density $\OmegaM =0.299\pm0.009$ \citep{bocquet14}.  The cluster $\fstar$ is in good agreement with this estimate of the universal average field value $\fstar=(0.95\pm0.12)\percent$ at $z=0$.  However, the average field $\fstar=0.22\pm0.003\percent$ (see Figure~\ref{fig:redshift_trends}) inferred from the SMF measurements at z=0.85--1.2 (vdB13) is significantly lower than the cluster \fstar.  The cluster or group \fstar\ may be altered over time through either the accretion of lower mass clusters or groups (higher \fstar) or through infall from the field (lower \fstar).  Presumably, these influences must combine to produce the transformation in \fstar\ from a lower mass cluster at $z=1$ to a higher mass clusters at $z=0$. We return to this discussion in Section~\ref{sec:discussion}.

We compare the high redshift SPT results to two other samples at high redshift: vdB14 and H13.  The virial masses for the majority of the vdB14 systems are below $3\times10^{14}\Msun$ and therefore lower mass than our SPT clusters.  The one remaining system in this mass range falls near the bottom of our distribution of $\fstar$.   The H13 sample shows stellar mass fractions that are in good agreement with ours.  We have three clusters in common; combining these measurements we determine that the differences are $1.11\sigma$, $0.69\sigma$ and $0.52\sigma$ for \Mstar, \Msz\ and \fstar\ measurements, respectively.  vdB14 express concern that the \Mstar\ estimated by the mass-to-light technique in H13 could possibly be overestimated by as much as a factor of 2.  While the largest difference with our sample is indeed with \Mstar, the level of agreement between the H13 results and our SED fitting results would suggest that the bias is likely smaller (about $10\pm23$\percent).


\subsection{ICM Mass Fraction \fgas}
\label{sec:ICM_literature}

The majority of the baryonic mass within clusters lies within the hot ICM.  The arithmetic mean of \fgas\ for the 14 SPT clusters is $0.1027 \pm 0.0073$, and the characteristic value at $z=0.9$ and $\Msz=6\times10^{14}\Msun$ is $0.096  \pm  0.005$.  A clear mass trend ($\fgas\propto\Msz^{0.43 \pm 0.13}$), significant at the $3.3\sigma$ level, is seen for SPT clusters.   This trend is steeper than (but statistically consistent with) the trends, $\beta_{\mathrm{ICM}} = 0.30 \pm 0.07$ and $0.26 \pm 0.03$, presented by Zha11 and GZ13,  but steeper at 2.1$\sigma$ than the result $\beta_{\mathrm{ICM}} = 0.15 \pm 0.03$ presented by  \citet{andreon10}.  The combined sample has a preferred mass trend $\beta_{\mathrm{ICM}} = 0.22 \pm 0.06$, which is 1.5$\sigma$ shallower than the SPT sample.

These results, extending to redshift $z=1.32$, show the clear tendency for \fgas\ to be suppressed in lower mass clusters--- first shown in studies of individual low redshift clusters \citep{david93} and later with a uniform analysis of a large sample of low redshift clusters \citep{mohr99}.

 The characteristic value of the combined sample at $z_{\mathrm{piv}}=0$ and $M_{\mathrm{piv}}=6\times10^{14}\Msun$ is $\fgas=0.112 \pm 0.0032$ (statistical only), which is higher than the $z=0.9$ SPT sample $\fgas=0.096 \pm 0.005$.  However, neither the SPT high redshift sample nor the combined sample exhibits evidence for redshift variation in \fgas\ with $\gamma_{\mathrm{ICM}} = -0.20  \pm  0.11\pm0.22$.  The impact of the halo mass 15\percent\ systematic uncertainty on the redshift trend introduces an additional systematic uncertainty of $\sigma_{\gamma_{\mathrm{ICM}}}=0.22$ which is larger than the statistical uncertainty.  This underscores the importance of using a homogeneous sample with consistently derived virial mass measurements.    

\subsection{Collapsed Baryon Fraction \fcold}
\label{sec:cold_literature}

The collapsed baryon fraction \fcold\ is the fraction of baryonic mass that has cooled to form stars that lie in galaxies and thereby reflects an integral of the star formation efficiency within the halo averaged over cosmic time \citep[e.g.][]{david92}.  As already noted, we make no attempt to include an estimate of the ICL contribution here.  The arithmetic mean \fcold\ of our fourteen clusters is $0.099 \pm  0.007$,
and the characteristic value at $z=0.9$ and $\Msz=6\times10^{14}\Msun$ is $0.107  \pm  0.011$ (statistical).  The SPT sample exhibits no evidence for either a mass or redshift trend.  The combined sample exhibits a 6.5$\sigma$ significant mass trend $\fcold\propto M^{-0.65\pm0.10}$
with the collapsed baryon fraction falling to lower values in high mass clusters and a characteristic value of $0.0859\pm0.0049$ (statistical) at $z=0$.  The redshift trend from the combined sample 
$\fcold\propto(1+z)^{0.39\pm0.15\pm0.16}$ is significant at 1.8$\sigma$ if we add the statistical and systematic uncertainties in quadrature.  

Note that in the case of \fcold\ the 15\percent\ systematic uncertainty in \Msz\ has no impact on the \fcold\ measurement, because a shift in \Rfiveoo\ has approximately the same impact on \Mstar\ and \Mgas,. However, because of the steep mass trend for \fcold\ ($\beta_{\mathrm{c}}\approx-0.6$), a shift in the virial mass of the high redshift sample impacts the best fit redshift trend, because this shift masquerades as a shift in \fcold\ of $\delta\fcold\sim-0.6\delta\Msz/\Msz$.
This impacts the estimate of the systematic uncertainty in $\gamma_{\mathrm{c}}$.  This process is also at work for the other fractions, but because their mass dependences are weaker, the impact is smaller.

In Figure~\ref{fig:mass_trends} we see that \fcold\ decreases with cluster mass, and the scatter about this trend (especially in the Zha11 sample) is less than in the case of \fstar.  This measure is interesting because the \Msz\ measurements come in only through defining the virial radius, and if the radial variation in \fgas\ and \fstar\ are mild, then \fcold\ has only a weak dependence on the virial mass estimates.  Thus in cases where the \Msz\ estimates exhibit large uncertainties, the \fcold\ can be an effective way of exploring trends in the mix of stars and ICM within clusters.

Our study indicates that over cosmic time the collapsed baryon fraction \fcold\ at fixed cluster halo mass falls.  This redshift trend is driven by the slight rise in the ICM mass fraction \fgas\ and slight fall in stellar mass fraction \fstar.     

\subsection{Baryon Fraction \fb}
\label{sec:baryon_literature}

The arithmetic mean of the baryon fraction for our SPT clusters is $0.114 \pm 0.008$ (statistical only), and the characteristic value at $z=0.9$ and $\Msz=6\times10^{14}\Msun$ is $0.107  \pm  0.006$ (statistical).  This is  lower than the characteristic values of the combined sample at $z=0$ of $0.1227  \pm  0.0035$ (statistical).  However, neither the SPT sample nor the combined sample ($\fb\propto(1+z)^{-0.17\pm0.11\pm0.22}$) provides clear evidence for a redshift trend.  The mass systematics between the low and high redshift samples introduce an uncertainty in the redshift trend parameter of $\sigma_{\gamma_{\mathrm{b}}}=0.22$, which is larger than the statistical uncertainty, implying that controlling mass systematics among the different samples is crucial.  The SPT sample exhibits a mass trend $\fb\propto\Msz^{0.39  \pm  0.13}$ that has 3$\sigma$ significance.  The combined sample exhibits a mass trend  $\fb\propto\Msz^{0.22  \pm  0.06}$, which is somewhat shallower and is significant at 3.6$\sigma$. 

%
%

\section{Discussion}
\label{sec:discussion}

Because our sample includes the highest redshift massive (\Msz$>3\times10^{14}$\Msun) clusters studied to date, our analysis is useful for constraining the redshift variation of the ICM and stellar mass components on cluster mass scales. 
While we do consider intrinsic scatter in fitting the observed properties within our sample, the sample does not provide meaningful constraints on this scatter;  thus, our results shed no light on assembly bias, which would link the baryon properties of individual clusters to the properties of the large scale environment within which they formed.
A joint analysis of the SPT sample and a comparison sample indicates that the cluster collapsed baryon fraction (accounting only for stars in galaxies) within \Rfiveoo\ is decreasing from 10.7\percent\ to 8.6\percent\ on the characteristic mass scale $M_{\mathrm{piv}}=6\times10^{14}\Msun$ since $z\approx0.9$; the redshift trend is significant at the 1.8$\sigma$ confidence level when accounting for a 15\percent\ virial mass systematic uncertainty between the literature and SPT samples.  Moreover our analysis indicates that this change is driven by a weak increase in the ICM fraction (\fgas\ changes from 9.6\percent\ to 11.2\percent) and a weak decrease in the stellar fraction from 1.1\percent\ to 1.0\percent\ over that same redshift range.  These same trends in \fgas\ and \fstar\ lead to a weak trend in the baryon fraction (from 10.7\percent\ to 12.3\percent) from $z=0.9$ to the present, a change that is only 0.7$\sigma$ significant given the systematic mass uncertainties between the high redshift and local comparison samples.

To build a physical picture it is important to take note of the mass trends in the stellar mass fraction $\fstar\propto M^{-0.37\pm0.09}$ that indicate that high mass clusters have \fstar\ values that lie below those of groups and that are comparable to or even higher than the field \fstar\ at $z=0$ (see also L03, vdB13).  The ICM mass fractions \fgas\ behave oppositely $\fgas\propto M^{0.22\pm0.06}$ \citep[see also][]{mohr99}, with groups having lower values than massive clusters, whose \fgas\ values are approaching but still lower than the universal baryon fraction.  These mass trends then give rise to the trend of falling collapsed baryon fraction \fcold\ with mass $\fcold\propto M^{-0.65\pm0.10}$.

Because of the clear mass trends and weak redshift trends in \fstar\ and \fgas, a simple merging scenario for halo formation, where the accretion of low mass (group-scale) halos is dominating the mass assembly of massive clusters, does not provide an adequate explanation of the observations.
In general, such a scenario would lead to \fstar\ that is approximately independent of cluster halo mass \citep{balogh08}.  The massive halos of today form from halos that were lower mass at higher redshift, so if these low mass subclusters had lower \fstar\ or higher \fgas\ at higher redshift, then the simple merger scenario could in principle be consistent with the data.  However, the weak redshift variation in these fractions at fixed halo mass that we estimate here for massive halos does not help to resolve the situation, because it indicates that \fstar\ and \fgas\ at fixed halo mass have changed only weakly over time;  if trends on the massive end are coupled with similar trends on the lower mass end, then the simple merging scenario must be flawed.  The conclusion that infall from the field and/or the inclusion of stripping processes that modify the apparent stellar fraction during the process of the growth of massive, cluster scale halos is inescapable.

Infall from the field likely plays a critical role in the growth of massive clusters.  Studies of the standard hierarchical structure formation scenario on the mass scales of interest here indicate that $\sim$40\percent\ of the cluster galaxies have previously been in lower mass group or cluster halos \citep{mcgee09} and that the rest infall from the field. In the case of \fstar, we have shown that the field has lower \fstar\ in comparison to massive clusters at redshift $z=0.9$ (Section~\ref{sec:smf}) and that it has comparable values of \fstar\ at  $z=0$ (see Figure~\ref{fig:redshift_trends}).  Through an appropriate mix of field and group accretion the \fstar\ values in massive clusters could in principle either increase or decrease with cosmic time.  Our results indicate that this mix of field and group accretion to build up the halos of the most massive clusters must produce halos with \fstar\ that are similar (at $\sim10$\percent\ level) up to (or weakly decreasing since) redshift $z\sim1$.

Add to this the likely stripping of stellar material from infalling galaxies during the accretion and relaxation process, and one has an additional mechanism to reduce the observed \fstar\ over cosmic time, because the ICL from these stripped stars is not included in the \fstar\ measurements here.  \citet{lin04b} suggested just such a mechanism to reconcile the falling \fstar\ with halo mass they observed in the local Universe.  They presented a toy model that suggested such a mechanism would have to lead to an ICL mass fraction that increases with halo mass and reaches high values of $\approx$40\percent\ of the stars in the central galaxy of the clusters.  Neither this trend nor ICL fractions at this high redshift have been observed in recent observational studies \citep{zibetti05,gonzalez13}.  Presumably, as massive clusters grow, a reduction of \fstar\ through both accretion from the field and stripping of stars from cluster galaxies is counterbalancing the increase of \fstar\ due to  accretion of lower mass subclusters.  Together these processes must transform high \fstar\ low mass clusters into lower \fstar\ high mass clusters.  Moreover, these processes must maintain a roughly constant \fstar\ at fixed halo mass over cosmic time on cluster mass scales.

A similar scenario of infall from the field and accretion of subclusters could explain the trends in \fgas\ as a function of halo mass and redshift.  In the case of \fgas, the field value, which is inferred by the \PLANCK\ measurement, is higher than that in the clusters at $z=0.9$ and remains so to $z=0$ (see Figure~\ref{fig:redshift_trends}).  Thus, given our observed weakly increasing cluster \fgas\ at fixed halo mass since $z\sim1$, the increases  in \fgas\ during cluster growth from infall from the field are compensating for the decreases in \fgas\ from accretion of subclusters.  These constraints, when coupled to a detailed hydrodynamical study, would presumably enable one to constrain processes such as early preheating as well as entropy injection from AGN residing in groups and clusters.

%
%

\section{Conclusions}
\label{sec:conclusion}

In this work we study the stellar mass function and baryon composition of 14 high redshift SZE-selected clusters between redshifts 0.572 and 1.32 that have a median mass \Msz\ of $6\times10^{14}$\Msun.  We estimate \fstar, \fgas, \fcold\ and \fb\ within \Rfiveoo\ (Table~\ref{tab:results}).  Our sample provides the highest redshift, uniformly selected sample to date for the study of the baryon content in massive clusters; our measurements together with low redshift measurements in the literature enable us to constrain the redshift variation of these quantities.  We summarize our results here.

\paragraph*{$\bullet$}
We examine the \MBCG--\Msz\ relation by combining our sample with the sample of H13 and vdB14 (Section~\ref{sec:bcg_mass}, Equation~\ref{eq:bcg}).  On the cluster mass scale of $6\times10^{14}\Msun$ the BCG stellar mass constitutes $0.12\pm0.01$\percent\ of the halo mass.  That fraction falls with cluster mass as $\Msz^{-0.58\pm0.07}$.  BCG stellar masses scatter about the best fit  \MBCG--\Msz\ relation with a characteristic value of 41\percent, a measure of the considerable variation in the BCG population.

\paragraph*{$\bullet$}
We measure the stacked SMF of these clusters and fit it to a Schechter function (Table~\ref{tab:smf}; Section~\ref{sec:smf}).  The characteristic mass is $\Mzero=10^{11.0\pm0.1}\Msun$, consistent with values derived in low mass clusters at high redshift (vdB14) and at low redshift \citep{vulcani13}.  Moreover, through comparison to constraints on the field SMF in the same redshift range (vdB13), we show that the number of galaxies with stellar mass above our threshold ($2.5\times10^{10}\Msun$) per unit total mass is higher in clusters than in the field by a factor of $1.65\pm0.20$. 

\bigskip
We take the measurements of the baryon composition in each of our clusters and fit to power law relations in redshift and mass (Equation~\ref{eq:fitting_form}).  We present best fit trends for the SPT sample and for a combined sample that includes several samples from the literature (Table~\ref{tab:power_law}).  In combining with external samples we homogenise the stellar mass measurements to the Chabrier IMF (Section~\ref{sec:IMF_correction}), we apply corrections for the differences in the virial mass estimates (Section~\ref{sec:mass_systematics}), we adopt a 15\percent\ (1$\sigma$) systematic virial mass uncertainty (Section~\ref{sec:fit_mass_systematic}), and we account for differences in the estimates of measurement uncertainties by solving for independent intrinsic scatter estimates for each subsample (Section~\ref{sec:fit_uncertainties}).  The key results are described below.

\paragraph*{$\bullet$}
The stellar mass fraction has a characteristic value $1.1\pm0.1$\percent\ (statistical) for clusters with mass $\Msz=6\times10^{14}\Msun$ at $z=0.9$ and $1.0\pm0.05$\percent\ (statistical) at $z=0$.  It falls with cluster halo mass $\fstar\propto M^{-0.37\pm0.09}$ and mildly decreases with cosmic time $\fstar\propto (1+z)^{0.26\pm0.16\pm0.08}$ with 1.45$\sigma$ significance, where the second component of the uncertainty represents the impact of the 15\percent\ systematic mass uncertainty between the low and high redshift samples.  
A similar result for the mass trend $\beta_{\star}\approx-0.26$ is also seen for low mass clusters and groups at $0.8\le z \le1.0$ \citep{balogh14}.
The mass trend and mild redshift trend indicate that the infall from subclusters (which would tend to increase \fstar) and infall from the field and stripping of stars from cluster galaxies (which would both tend to decrease the observed \fstar) must combine to enable the transformation of \fstar\ from low mass clusters into that of higher mass clusters having similar \fstar\ over the redshift range $0<z<1$.  Numerical simulations suggest that approximately 40\percent\ of cluster galaxies have been accreted as members of subclusters, and the remainder from the field \citep{mcgee09}, but additional study is warranted to test whether the observed trends in \fstar\  (now constrained both as a function of mass and of redshift) can be reproduced by current structure formation scenarios.

\paragraph*{$\bullet$}
The ICM mass fraction has a characteristic value in clusters with mass $\Msz=6\times10^{14}\Msun$ of $9.6\pm0.5$\percent\ (statistical) at $z=0.9$ and $11.2\pm0.32$\percent\ (statistical) at $z=0$.  It rises with cluster halo mass $\fgas\propto M^{0.22\pm0.06}$ and evolves weakly with redshift at fixed halo mass as $\fgas\propto (1+z)^{-0.20\pm0.11\pm0.22}$, where the $0.22$ is due to the 15\% systematic mass uncertainty between the low and high redshift samples.  The trend of increasing \fgas\ with mass has been previously observed \citep{mohr99} and can be explained through entropy injection through early preheating or from cluster AGN.  A weakly varying \fgas\ with cosmic time could be explained by infall from the field, where \fgas\ is larger than that in clusters at $z=0.9$ and $z=0$ (see Figure~\ref{fig:redshift_trends}).  Hydrodyamical studies of this scenario are needed.

\paragraph*{$\bullet$}
The collapsed baryon fraction determines the fraction of the baryonic component that has cooled to form stars.  It is the ratio of the stellar mass to the ICM plus stellar mass.  The characteristic value at cluster masses $\Msz=6\times10^{14}\Msun$ is $10.7\pm0.1$\percent\  (statistical) at $z=0.9$ and $8.6\pm0.5$\percent\ (statistical) at $z=0$.  It falls with halo mass as $\fcold\propto M^{-0.65\pm0.10}$, indicating with 6.5$\sigma$ significance that a smaller fraction of halo baryons is in the form of stars in the most massive halos.  The redshift trend is $\fcold\propto (1+z)^{0.39\pm0.15\pm0.16}$, where the second uncertainty is due to the 15\percent\ systematic mass uncertainty between the low and high redshift samples.  Thus, there is $\approx1.8\sigma$ evidence that the collapsed  baryon fraction is falling with cosmic time, and this is driven by the weak trends of rising \fgas\ and falling \fstar\ presented above.  

\paragraph*{$\bullet$}
The baryon fraction \fb\ is the fraction of the halo mass that is in ICM and stars.   The characteristic value at cluster mass $\Msz=6\times10^{14}\Msun$ is $10.7\pm0.6$\percent\ (statistical) at $z=0.9$ and $12.3\pm0.4$\percent\ (statistical) at $z=0$.  It rises with halo mass as $\fb\propto M^{0.22\pm0.06}$, and this 3.7$\sigma$ mass trend is affected both by the increase in \fgas\ and the decrease in \fstar\ with cluster mass.   The evidence for redshift variation at fixed halo mass is weak $\fb\propto (1+z)^{-0.17\pm0.11\pm0.22}$, where the second uncertainty is due to the 15\% systematic mass uncertainty between the low and high redshift samples.  If the two uncertainties are added in quadrature, there is no significant evidence (0.7$\sigma$) that the baryon fraction is evolving.

As already discussed in Section~\ref{sec:discussion}, these mass and redshift trends in baryon quantities are not consistent with a simple hierarchical structure formation merger model where massive clusters form solely through the accretion of lower mass clusters and groups.  Significant accretion of galaxies and ICM from the field must also occur, and this accretion together with infall of subclusters can likely explain the weak variation (at fixed cluster halo mass) in \fstar\ and in \fgas\ over cosmic time.  Additionally, the loss of stellar mass from galaxies through stripping is an additional mechanism that would allow for \fstar\ to fall as low mass clusters grow to higher mass.

This analysis of the first homogeneously selected high mass cluster sample extending to high redshift allows for interesting initial constraints on the redshift trends in the baryon content; however, these trends are dependent to some extent on the adopted systematic virial mass uncertainty between the low and high redshift samples.  If the systematic virial mass uncertainty is 15\% there are no statistically significant redshift trends.  Higher virial mass systematic uncertainties would further reduce the significance of trends.  A reduction of the 15\percent\ systematic virial mass uncertainty to 10\percent\ or 5\percent\ would result in a fractional reduction (to $2/3$ or $1/3$, respectively) for the redshift trend systematic uncertainties $\gamma^{\mathrm{sys}}_{\mathrm{obs}}$.  In the case of a 5\percent\ systematic virial mass uncertainty, the significance of the redshift trends for \fstar\ (\fgas, \fcold, \fb) would increase from $1.12\sigma$ ($0.81\sigma$, $1.6\sigma$, $0.69\sigma$) to $1.27\sigma$ ($1.53\sigma$, $2.2\sigma$, $1.28\sigma$).  It is clear that what is needed is a systematic study of a large, homogeneously selected cluster sample with high quality mass estimates that spans a broad redshift range. 

\vskip0.5cm
%
%

We acknowledge the support by the DFG Cluster of Excellence ``Origin and Structure of the Universe'' and the Transregio program TR33 ``The Dark Universe''. The calculations have been carried out on the computing facilities of the Computational Center for Particle and Astrophysics (C2PAP) and of the Leibniz Supercomputer Center (LRZ).  BB is supported by the Fermi Research Alliance, LLC under Contract No. De-AC02-07CH11359 with the United States Department of Energy.  BS acknowledges the support of the NSF grants at Harvard and SAO (AST-1009012, AST-1009649 and MRI- 0723073). TS acknowledges the support from the German Federal Ministry of Economics and Technology (BMWi) provided through DLR under project 50 OR 1210.  The South Pole Telescope is supported by the National Science Foundation through grant ANT-0638937.  Partial support is also provided by the NSF Physics Frontier Center grant PHY-0114422 to the Kavli Institute of Cosmological Physics at the University of Chicago, the Kavli Foundation and the Gordon and Betty Moore Foundation. 

Optical imaging data from the VLT programs 088.A-0889 and 089.A-0824, HST imaging data from programs C18-12246 and C19-12447, and Spitzer Space Telescope imaging from programs 60099, 70053 and 80012 enable the SED fitting in this analysis.  X-ray data obtained with Chandra X-ray Observatory programs and XMM-{\it Newton} Observatory program 067501 enable the ICM mass measurements.  The SPT survey program SPT-SZ enabled the discovery of these high redshift clusters and subsequent analyses have enabled virial mass estimates of these systems.  Optical spectroscopic data from VLT programs 086.A-0741 and 286.A-5021 and Gemini program GS-2009B-Q-16, GS-2011A-C-3, and GS-2011B-C-6 were included in this work.  Additional spectroscopic data were obtained with the 6.5~m Magellan Telescopes.

Facilities:  South Pole Telescope, \Spitzer/IRAC, VLT: Antu (FORS2), \HST/ACS, \CHANDRA, \XMMNEWTON, Magellan

%
%

\bibliographystyle{mn2e}
\bibliography{spt}

%
%

\appendix
\section{Performance of SED fitting}
\label{sec:specz}

\begin{figure}
\centering
\includegraphics[scale=0.5]{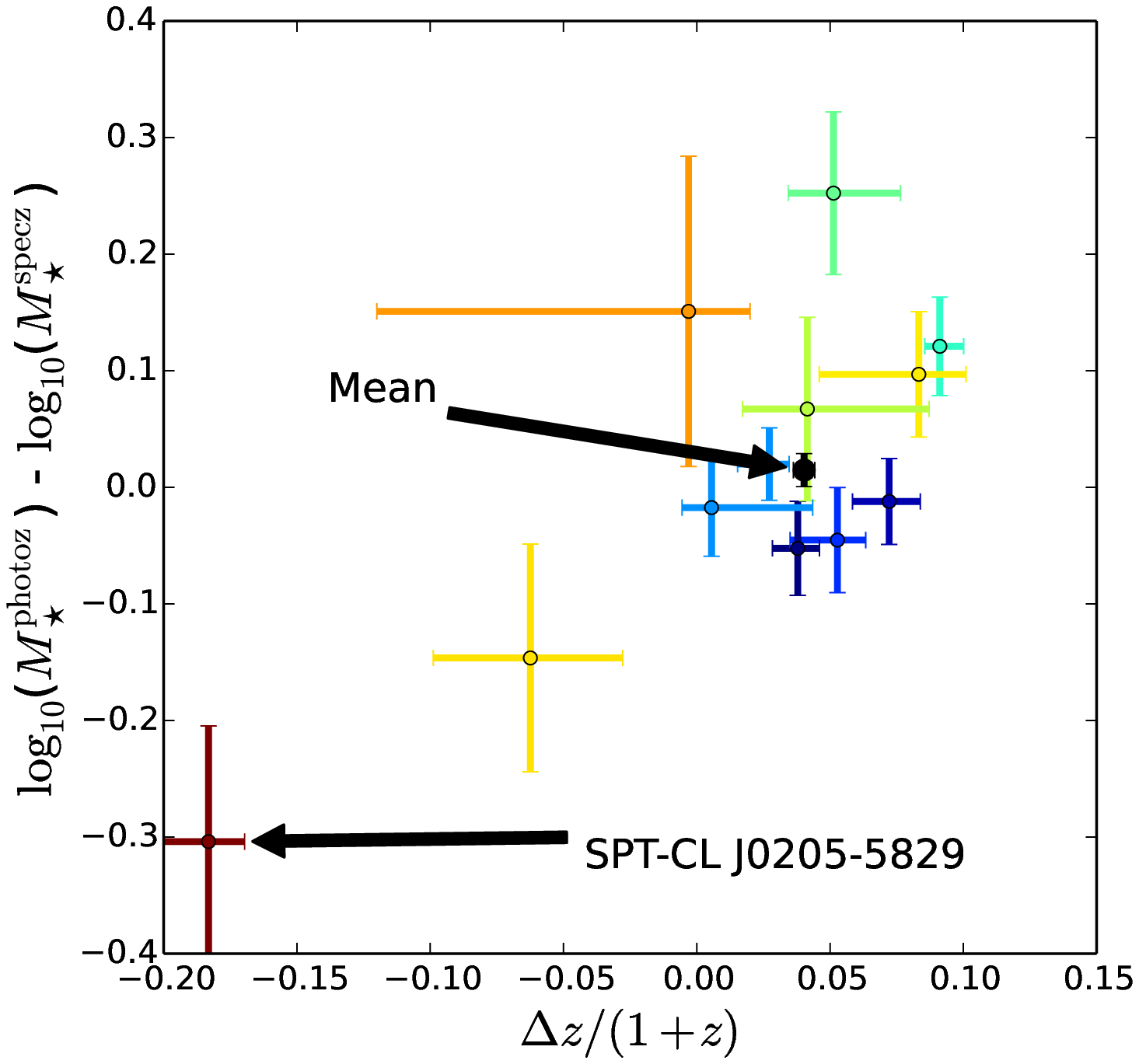}
\caption{A test of SED fitting using galaxies with spectroscopic redshifts. The x-axis is the normalized difference of photo-z and spec-z and the y-axis is the resulting stellar mass difference averaged on a per-cluster basis. The color code from blue to red indicates the clusters from the lowest to the highest redshift. SPT-CL~J0205-5829 at $z=1.32$ is marked as it has the largest mass difference. The black dot indicates the mean of ($\log_{10}M^{\mathrm{ photoz}}_{\star} - \log_{10}M^{\mathrm{ specz}}_{\star}$) and $\Delta z / (1+z)$ of the ensemble of clusters.}
\label{fig:dmdz}
\end{figure}

With the published spectroscopic sample for SPT clusters \citep{sifon13,ruel14}, we are able to quantify how the uncertainty of the photo-z impacts on the stellar mass estimates based on the SED fit using the six band photometry (\BHigh, \Fband, \IBess, \ZGunn, \Cha, \Chb).  We cross-match our photometry identified sample with the galaxy sample in \cite{ruel14} and repeat the whole SED fit analysis with the redshift fixed to the measured spectroscopic redshift. We show the comparison in Figure~\ref{fig:dmdz}. The photo-z performance is estimated as the mean $\Delta z / (1+z) \equiv (z_{\mathrm{ photo}} - z_{\mathrm{ spec}}) / (1+z_{\mathrm{ spec}})$ to be $0.037 \pm 0.0083$.  The difference of the stellar mass estimates ($\log_{10}M^{\mathrm{ photoz}}_{\star} - \log_{10}M^{\mathrm{ specz}}_{\star}$) when using $z_{\mathrm{ photo}}$ and $z_{\mathrm{ spec}}$ is at the level of $\lessapprox0.2$ with a mean $\approx0.03$. Except for the highest redshift cluster (SPT-CL~J0205-5829 at $z=1.32$), which has only 5 spectroscopic redshifts available for the cluster members, the SED fitting using our six band photometry returns unbiased estimates of the stellar masses. 
\newline

\section{Tests of Statistical Background Correction}
\label{sec:background_test}

To test the COSMOS background, we extract the local background information from our SPT dataset, applying a correction for the cluster galaxy contamination. We extract the corrected local background between 1.2\Rfiveoo\ and 2.5\Rfiveoo\ for each cluster.  We correct for cluster contamination by assuming that the cluster galaxies are distributed as an NFW model with concentration of $c^{\mathrm{gal}}_{500}=1.9$ \citep[][Hennig in prep]{lin04a}, and the Stellar Mass Function (SMF) and the Magnitude Distribution (MD) are the same for the region within the cluster \Rfiveoo\ and for the cluster population that is contaminating the background region. Together with the area extracted for the region within \Rfiveoo\ and the local background, we solve for the the surface number densities of the SMF and MD using the corrected local background for each cluster. The SMF and MD derived using the corrected local background are noisy for each individual cluster, especially for the lower redshift clusters where the area available for the local background is typically less than 5~arcmin$^2$.  We combine 9 of the 14 independent estimates (those with background area larger than 8~arcmin$^2$) to create an average local background estimate.  In averaging, we use the area weighted average of the individual background estimates so that clusters with greater area (but not necessarily higher number density) receive higher weight.  

\begin{figure}
\centering
\includegraphics[scale=0.44]{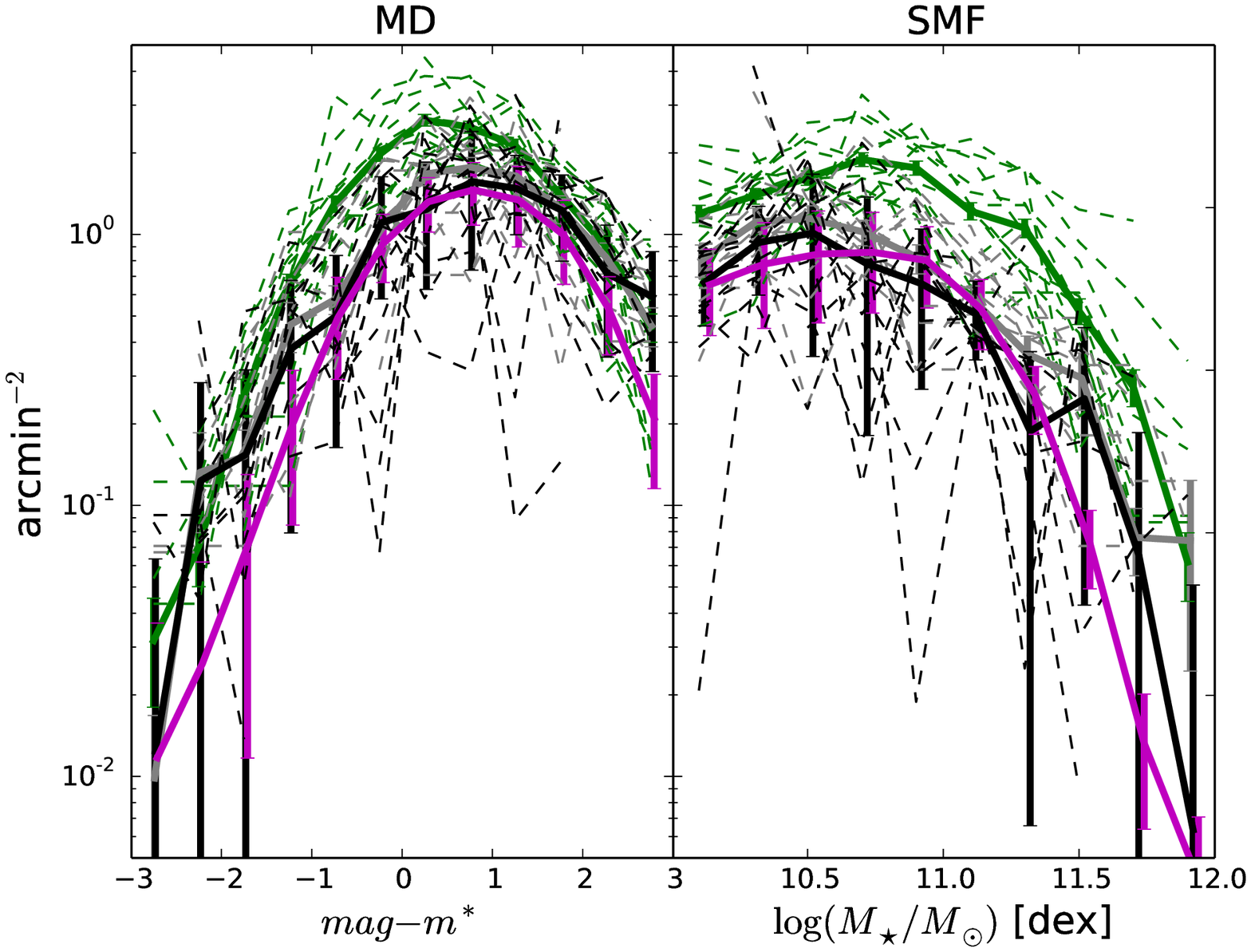}
\caption{The magnitude distribution (MD- left) and stellar mass function (SMF- right) for the full population of galaxies in the SPT clusters.  We show the cluster + background estimates from within \Rfiveoo\ (green), the uncorrected local background (grey), the corrected local background (black), and the background estimated from COSMOS (magenta). The SMFs are derived using SED fitting of six band photometry. The dashed lines indicate the results for individual clusters and the heavy-solid lines are the averages over all clusters.  The COSMOS and local, contamination-corrected background estimates are in good agreement.  We adopt the COSMOS background correction in this work.}
\label{fig:smf_md}
\end{figure}

Figure~\ref{fig:smf_md} contains a comparison of the COSMOS and local background estimates for the SMF (right panels) and MD (left panels).  The corrected local background estimates (black) for the SMF and MD are in a good agreement with the COSMOS backgrounds (magenta).  There is poorer agreement on the bright (massive) end with the tendency that the local background is slightly higher than COSMOS.  The cluster plus background SMF and MDs extracted from within \Rfiveoo\ (green) show significant overdensities with respect to the background estimates.  In both the case of the local background estimates (black) and the cluster plus background estimates (green), the individual cluster results are shown with dashed lines and the thick solid lines represent the ensemble average. 

On the other hand, the corrected local background for the SMF and MD for the red population is generally lower than the COSMOS estimates. This suggests we are overcorrecting the local backgrounds for cluster contamination in the case of the red population, and this is to be expected given that we do not have the right filter combinations (blue band containing 4000~\AA\ break and one band redward of the break) for the half of our sample that lies at $z>0.9$.  For these reasons we do not present any analyses of the red sequence selected subpopulation in this paper.

We compare the differences between the cumulative stellar mass estimates for the full population when using the two different background corrections.  We fit a simple linear relation $M_{\mathrm{local}} = 10^{x}\times M_{\mathrm{COSMOS}}$, allowing the normalization $10^x$ to float, where $M_{\mathrm{local}}$ and $M_{\mathrm{COSMOS}}$ are the mass estimations for using the local and COSMOS backgrounds, respectively.  The resulting best-fit $x$ is $-0.018 \pm 0.005$ ($0.045 \pm 0.012$) for the cluster (background) stellar mass estimation.  That is, using the COSMOS background results in $\sim$4\percent\ higher  stellar mass estimates for the cluster and $\sim$10\percent\ lower mass estimates in the background as compared to those using the corrected local background.

%
%

\clearpage
\begin{landscape}
\begin{table}
\caption{Measured quantities for the SPT cluster sample: Columns contain the cluster name, spectroscopic redshift, \Msz\ estimated from the SZE signature, \Rfiveoo\ inferred from the given \Msz\ and redshift, ICM mass \Mgas, the BCG mass \MBCG, the total stellar mass \Mstar, the stellar mass fraction \fstar, the collapsed baryon fraction \fcold, the  baryon fraction \fb, the ICM mass fractions \fgas\ and the stellar-mass-to-light ratios $\Upsilon$ (rms in the parenthesis) in \Cha\ band in the observed frame.}
\begin{tabular}{cccccccccccc}
\hline \hline \\
 & &\Msz &\Rfiveoo &\Mgas &\MBCG &\Mstar & \fstar &\fcold &\fb &\fgas\ & $\Upsilon$ \\ [6pt]
Cluster & Redshift & $[10^{14}\Msun]$ & [Mpc]  & $[10^{13}\Msun]$ & $[10^{11}\Msun]$ & $[10^{12}\Msun]$ & [\percent] &  [\percent] &  [\percent] &  [\percent]  & [$\frac{\Msun}{L_{\odot}}$] \\
\hline
     SPT-CL~J0000-5748      &  0.702      &  4.35 $\pm$ 1.16      &  0.90      &  4.33 $\pm$ 0.73      &  13.49$^{ +1.18 }_{ -1.06 }$   &  6.34 $\pm$ 2.22      &  1.46 $\pm$ 0.64      &  12.76 $\pm$ 4.33      &  11.42 $\pm$ 3.51      &  9.96 $\pm$ 3.14      &  0.37 (  0.146  ) \\    [6pt]       
     SPT-CL~J0102-4915      &  0.870      &  15.75 $\pm$ 3.22      &  1.30      &  25.51 $\pm$ 0.98      &  9.23$^{ +0.82 }_{ -0.73 }$   &  10.29 $\pm$ 1.92      &  0.65 $\pm$ 0.18      &  3.88 $\pm$ 0.71      &  16.85 $\pm$ 3.50      &  16.19 $\pm$ 3.37      &  0.32 (  0.196  ) \\    [6pt]       
     SPT-CL~J0205-5829      &  1.320      &  5.65 $\pm$ 1.14      &  0.78      &  5.26 $\pm$ 1.13      &  3.95$^{ +0.92 }_{ -1.02 }$   &  5.01 $\pm$ 0.96      &  0.89 $\pm$ 0.25      &  8.70 $\pm$ 2.28      &  10.20 $\pm$ 2.87      &  9.31 $\pm$ 2.74      &  0.42 (  0.158  ) \\    [6pt]       
     SPT-CL~J0533-5005      &  0.881      &  4.24 $\pm$ 1.13      &  0.83      &  2.70 $\pm$ 0.51      &  5.89$^{ +0.56 }_{ -0.47 }$   &  2.95 $\pm$ 0.98      &  0.70 $\pm$ 0.30      &  9.85 $\pm$ 3.39      &  7.06 $\pm$ 2.24      &  6.36 $\pm$ 2.07      &  0.33 (  0.221  )\\    [6pt]       
     SPT-CL~J0546-5345      &  1.067      &  5.48 $\pm$ 1.16      &  0.85      &  7.05 $\pm$ 1.23      &  17.41$^{ +1.86 }_{ -4.57 }$   &  14.07 $\pm$ 2.19      &  2.57 $\pm$ 0.67      &  16.63 $\pm$ 3.24      &  15.44 $\pm$ 3.99      &  12.87 $\pm$ 3.53      &  0.38 (  0.208  ) \\    [6pt]       
     SPT-CL~J0559-5249      &  0.609      &  7.16 $\pm$ 1.44      &  1.11      &  8.36 $\pm$ 0.55      &  4.83$^{ +0.41 }_{ -0.38 }$   &  8.10 $\pm$ 1.64      &  1.13 $\pm$ 0.32      &  8.83 $\pm$ 1.71      &  12.80 $\pm$ 2.70      &  11.67 $\pm$ 2.47      &  0.30 (  0.171  ) \\    [6pt]       
     SPT-CL~J0615-5746      &  0.972      &  11.75 $\pm$ 2.35      &  1.13      &  13.60 $\pm$ 2.25      &  14.39$^{ +1.75 }_{ -4.61 }$   &  14.70 $\pm$ 2.45      &  1.25 $\pm$ 0.33      &  9.75 $\pm$ 2.07      &  12.83 $\pm$ 3.21      &  11.57 $\pm$ 3.00      &  0.36 (  0.186  ) \\    [6pt]       
     SPT-CL~J2040-5726      &  0.930      &  4.10 $\pm$ 0.97      &  0.81      &  4.25 $\pm$ 0.95      &  5.44$^{ +0.49 }_{ -0.45 }$   &  3.44 $\pm$ 0.93      &  0.84 $\pm$ 0.30      &  7.50 $\pm$ 2.43      &  11.22 $\pm$ 3.53      &  10.38 $\pm$ 3.38      &  0.28 (  0.146  ) \\    [6pt]       
     SPT-CL~J2106-5844      &  1.132      &  9.35 $\pm$ 1.84      &  0.99      &  11.68 $\pm$ 1.43      &  9.96$^{ +0.87 }_{ -0.79 }$   &  14.06 $\pm$ 2.01      &  1.50 $\pm$ 0.37      &  10.75 $\pm$ 1.80      &  13.99 $\pm$ 3.16      &  12.49 $\pm$ 2.89      &  0.42 (  0.171  ) \\    [6pt]       
     SPT-CL~J2331-5051      &  0.576      &  6.45 $\pm$ 1.34      &  1.08      &  6.07 $\pm$ 0.83      &  3.03$^{ +2.67 }_{ -0.66 }$   &  2.78 $\pm$ 1.31      &  0.43 $\pm$ 0.22      &  4.38 $\pm$ 2.06      &  9.84 $\pm$ 2.42      &  9.41 $\pm$ 2.34      &  0.37 (  0.197  )\\    [6pt]       
     SPT-CL~J2337-5942      &  0.775      &  9.44 $\pm$ 1.83      &  1.14      &  8.52 $\pm$ 0.79      &  10.33$^{ +1.00 }_{ -0.80 }$   &  11.24 $\pm$ 2.14      &  1.19 $\pm$ 0.32      &  11.66 $\pm$ 2.18      &  10.21 $\pm$ 2.16      &  9.02 $\pm$ 1.94      &  0.40 (  0.156  ) \\    [6pt]       
     SPT-CL~J2341-5119      &  1.003      &  6.59 $\pm$ 1.31      &  0.92      &  6.85 $\pm$ 1.00      &  9.41$^{ +1.62 }_{ -1.84 }$   &  7.12 $\pm$ 1.60      &  1.08 $\pm$ 0.32      &  9.42 $\pm$ 2.29      &  11.47 $\pm$ 2.75      &  10.39 $\pm$ 2.57      &  0.34 (  0.183  ) \\    [6pt]       
     SPT-CL~J2342-5411      &  1.075      &  4.43 $\pm$ 1.07      &  0.79      &  3.15 $\pm$ 0.64      &  8.25$^{ +0.74 }_{ -0.67 }$   &  5.03 $\pm$ 1.17      &  1.14 $\pm$ 0.38      &  13.77 $\pm$ 3.66      &  8.25 $\pm$ 2.47      &  7.11 $\pm$ 2.24      &  0.39 (  0.167  ) \\    [6pt]       
     SPT-CL~J2359-5009      &  0.775      &  4.98 $\pm$ 1.16      &  0.92      &  3.50 $\pm$ 0.34      &  6.92$^{ +0.58 }_{ -0.53 }$   &  4.80 $\pm$ 1.31      &  0.96 $\pm$ 0.35      &  12.06 $\pm$ 3.07      &  8.00 $\pm$ 2.00      &  7.03 $\pm$ 1.77      &  0.43 (  0.182  ) \\    [6pt]       
\hline
\label{tab:results}
\end{tabular}
\end{table}
\clearpage
\end{landscape}

\end{document}